\documentclass[12pt,a4paper,DIV12]{scrartcl}
\usepackage{subfigure}
\usepackage{jheppub}
\pdfoutput=1
\makeatletter
\DeclareOldFontCommand{\rm}{\normalfont\rmfamily}{\mathrm}
\DeclareOldFontCommand{\sf}{\normalfont\sffamily}{\mathsf}
\DeclareOldFontCommand{\tt}{\normalfont\ttfamily}{\mathtt}
\DeclareOldFontCommand{\bf}{\normalfont\bfseries}{\mathbf}
\DeclareOldFontCommand{\it}{\normalfont\itshape}{\mathit}
\DeclareOldFontCommand{\sl}{\normalfont\slshape}{\@nomath\sl}
\DeclareOldFontCommand{\sc}{\normalfont\scshape}{\@nomath\sc}
\makeatother
\usepackage[utf8]{inputenc}     
\usepackage[T1]{fontenc}
\usepackage{lmodern}
\usepackage[british]{babel}
\usepackage{amsmath,bm}
\usepackage{amssymb}
\usepackage{amsfonts}
\usepackage{color}
\usepackage{float}
\usepackage{appendix}
\newcommand{\Ii}{\ensuremath{\mathrm{i}\hspace{1pt}}}

\newcommand{\tr}{\ensuremath{\mathrm{tr}}}

\newcommand{\VR}{\ensuremath{V^{(n)}_{\textrm{eff}}}}

\newcommand{\arxiv}[1]{arXiv:\,\href{http://arxiv.org/abs/#1}{{\tt #1}}}
\newcommand{\arxivv}[2]{[arXiv:\,\href{http://arxiv.org/abs/#1}{{\tt #1}}[{\tt #2}]]}
\usepackage{slashed}
\usepackage{amsmath}

\title{Adiabatic  continuity and confinement in  supersymmetric Yang-Mills theory on the lattice}
\author[a]{Georg Bergner,}
\author[b]{Stefano Piemonte,}
\author[c]{Mithat  \"Unsal,}
\affiliation[a]{Friedrich-Schiller-University Jena, Institute of Theoretical Physics,
Max-Wien-Platz 1, D-07743 Jena, Germany}
\affiliation[b]{University of Regensburg, Institute for Theoretical Physics,
Universit\"atsstr.~31, D-93040 Regensburg, Germany}
\affiliation[c]{Department of Physics, North Carolina State University, Raleigh, NC 27695, USA}
\emailAdd{georg.bergner@uni-jena.de}
\emailAdd{stefano.piemonte@ur.de}
\emailAdd{unsal.mithat@gmail.com}

\abstract{ 
This work is a step towards merging the ideas that arise from   semi-classical methods in continuum QFT  
   with analytic/numerical  lattice field  theory. 
 In this context, we consider  Yang-Mills theories coupled to fermions in the adjoint representation. These theories   have the remarkable property that 
 confinement and discrete chiral symmetry breaking  can persist  at weak coupling on 
 small  (non-thermal) $\mathbb R^3 \times S^1$. 
This work presents a lattice investigation of Yang-Mills with one-adjoint Majorana fermion, $\mathcal N=1$ super Yang-Mills, and 
     opens the prospect to understand a number of  non-perturbative phenomena, such as 
     the mechanism of confinement,  mass gap,  chiral  and center  symmetry realizations   both in lattice and continuum  analytically.  
  We study the  compactification of this theory on the lattice with periodic and thermal boundary conditions. We provide numerical evidence for the conjectured absence of the phase transitions with periodic boundary conditions for sufficiently light lattice fermions (stability of center-symmetry),       suppression of the chiral transition, and also provide a diagnostic for abelian vs. non-abelian confinement,   based on per-site Polyakov loop eigenvalue distribution functions.
    In numerical and perturbative investigations we identify  the role of the lattice artefacts that become relevant in the very small radius regime, and resolve some puzzles in the naive comparison between continuum and lattice. 
}

\begin{document}

\maketitle
\flushbottom
\section{Introduction}

Confinement is a  feature of strong interactions emerging in the long distance physics of certain non-Abelian gauge theories. The effective low energy degrees of freedom of these theories are colorless bound states which cannot be described in simple terms by the fundamental fields.  Non-perturbative features of the strongly interacting regime like the bound state particle spectrum are consequently still beyond any analytical understanding.  Despite the success of the numerical lattice simulations in reproducing the observed hadron masses, an understanding of the nature of confinement is still missing on $\mathbb R^{4} $.  Perturbation theory can provide a reliable approximation of scattering processes only at very high energy, but it cannot explain the structure and the properties of low energy states.

An interesting approach for a better analytical understanding is based on a  controlled semiclassical analysis that tries to identify the most relevant field contributions  as in the Polyakov model on $\mathbb R^3$ \cite{Polyakov:1976fu}.   However, despite the important  success of the Polyakov model, these ideas remained  dormant in QCD context, mainly due to  two reasons: One is that it was often believed that the Polyakov mechanism was a manifestly 3d mechanism, and the other is that in theories with exactly massless fermions, the  Polyakov mechanism would not work \cite{Affleck:1982as}.      About ten years ago, new ideas and techniques in semi-classics started to emerge and  the utility of  (justified) semi-classics into non-perturbative problems in  QCD and other vector-like and chiral gauge theories  has been realized  \cite{Unsal:2007vu, Unsal:2007jx,Unsal:2008ch, Shifman:2008ja}.  For related earlier work, see   \cite{Seiberg:1996nz, Lee:1997vp, Kraan:1998pm, Davies:1999uw}.

More recently theories with matter fields in the adjoint representation of the gauge group and certain supersymmetric theories became a valuable subject of these investigations \cite{Argyres:2012ka,Poppitz:2013zqa} using tools of resurgence,  and Picard-Lefschetz theory.  The QFT application of resurgence   may potentially provide a rigorous non-perturbative definition of path integral in continuum QFT.  At least, in certain quantum mechanichal context with instantons,  it is proven that  path integral can be decoded into a full semi-classical resurgent expansion \cite{Dunne:2013ada}, and in certain asymptotically free  2d QFTs, such as 2d ${\mathbb {CP}}^{N-1} $  and principle chiral  models,  ideas related to resurgence have been fruitful to partially resolve the renormalon problem \cite{Dunne:2016nmc, Dunne:2012ae, Cherman:2014ofa}.

The semiclassical analysis corresponds to a sum of the expansions around the dominant field configurations in a regime where 
weak coupling methods are reliable. (See  \cite{Dunne:2016nmc}   for a recent review.) The inclusion of non-trivial configurations should overcome the limitation of standard perturbation theory by incorporating exponentially small non-perturbative  effects of the form $e^{-A/(g^2N)}$ where $A$ is pure number and $(g^2N)$ is the 't Hooft coupling.\footnote{\;Note that this type of non-perturbative effect is exponentially larger than the usual  4d instanton effects in gauge theories, which are of the form  $e^{-A/g^2}$. } 

 If one space-time dimension is thermally compactified and the size of the compact dimension $R$ is smaller than the strong length  scale $\Lambda^{-1}$,  the theory becomes weakly coupled due to asymptotic freedom. 
In this regime,      QCD and Yang-Mills theories are in the deconfined phase as can be shown by studying 
Gross-Pisarski-Yaffe  (GPY) potential  \cite{Gross:1980br}.    Consequently this  region  is disconnected by a phase transition (or crossover) from the large $R$ low energy confined phase that one aims  to understand.  Furthermore, the dynamics of the theory at distances much larger than $R$ is governed by 3d pure Yang-Mills, which develops the strong (magnetic)  scale of its own.  

The situation is different in a gauge theory where the fermion fields are in the adjoint representation and fulfill periodic boundary conditions in the compact direction.  Periodic boundary condition in path integral formalism map to 
\begin{align}
\tilde Z(R, m) =  \; \tr [e^{-R \hat{H}(m)} (-1)^F]  
\label{tpf}
\end{align}
in the operator formalism, where $\hat{H}(m)$ is the Hamilton operator and $m$ is the fermion mass.   This trace is a graded state sum which assigns an over-all (+1) sign to boson states and (-1) to fermion states.  Note that \eqref{tpf} for $N_f=1$ and  $m=0$ is the supersymmetric Witten index \cite{Witten:1982df}. For general $N_f$ or $m \neq 0$, it is a  (non-thermal) twisted partition function which probes the phase structure of the theory as a function of compactification radius \cite{Unsal:2007jx,  Poppitz:2012sw}

 If we consider the one-loop GPY potential with this partition function, we observe that the center-destabilizing  gauge boson induced   potential is overwhelmed   by the    center stabilizing    contributions from sufficiently light fermion fields \cite{Hosotani:1983xw, Kovtun:2007py}. In this case, the small $R$ regime is in a center symmetric  phase on $\mathbb R^3 \times S^1$ which is non-perturbatively  calculable by semiclassical methods. The theory exhibits semiclassical  magnetic bion  mechanism of confinement, a non-perturbative mass gap for gauge  fluctuations, and, if it exists in the theory, a discrete chiral symmetry breaking. 

The small $R$ regime on $\mathbb R^3 \times S^1$ provides a controlled semi-classical approximation in sharp distinction from the so-called dilute instanton gas picture on $\mathbb R^4 $ which is an uncontrolled approximation, see section "the uses of instantons" in \cite{coleman}  and \cite{Schafer:1996wv} for the limitations of this approach. The calculable semi-classical regime might in certain cases be smoothly connected to the confined strongly coupled regime at large $R$  \cite{Unsal:2007jx,   Unsal:2008ch, Shifman:2008ja,  Anber:2018iof}. This is the notion of {\it adiabatic continuity } which posits that a semi-classical calculable regime, under appropriate conditions,  may be smoothly connected to a strong coupling regime without any intervening  phase transitions.

 An exact cancellation between fermion and boson perturbative contributions occurs in the particular case of the $\mathcal{N}=1$ Supersymmetric Yang-Mills theories (SYM)  or $N_f=1$ QCD(adj) \cite{Davies:1999uw, Anber:2014lba}. 
Both the center-stability as well as the confinement mechanisms of compactified SYM are  solely due to non-trivial semiclassical contributions, namely neutral and magnetic bions.  The phase structure of gauge theories with $N_f$ adjoint Majorana fermions ($N_f$-flavour QCD(adj)) opens therefore a useful perspective for the understanding of confinement \cite{Unsal:2007jx, Argyres:2012ka}, see also \cite{Kouno:2013mma,Kashiwa:2013rmg,Anber:2013doa}.   

An a-priori non-perturbative regularization is however required  in order to prove a continuous connection of the small $R$ regime to the strongly coupled  large $R$  confined  phase and to verify the absence of intermediate phase transitions. Numerical lattice simulations are an ideal first principles method. 

The first observation of the absence of deconfinement in compactified supersymmetric Yang-Mills theory has been presented in an earlier publication \cite{Bergner:2014dua}, we are investigating the phases of the theory more closely in this work. In order to understand the influence of the lattice discretization, we first compare the perturbative analysis in the continuum and on the lattice. It turns out that the discretization of the fermion action on the lattice is of particular importance in the small $R$ regime. Due to lattice artefacts, the exact cancellation between fermion and boson contributions can only be achieved in the continuum limit. Nevertheless we show that qualitative features of the semiclassical predictions are reproduced on the lattice. In contrast to our earlier investigations we concentrate here on a fixed scale approach which avoids additional complications introduced by a variation of the lattice spacing. Part of the work is done with a clover improved Wilson fermion action in order to reduce the discretization effects. We consider the order parameters for the chiral and the deconfinement transition. The primary focus is on supersymmetric Yang-Mills theory, but we have also investigated the $N_f$ dependence in a numerical study of $N_f=2$ QCD(adj).

In the next two chapters we introduce the general lattice formulations and the order parameters for the investigations of the phase transitions. In Section~\ref{sec:veff} we provide a detailed discussion of the perturbative effective potential of the Polyakov loop on the lattice and in the continuum. Based on the analysis of the effective potential, we derive predictions of the phase diagram of QCD(adj) in the weak coupling regime and conjectures about the general phase diagram in Section~\ref{phase}. After these theoretical considerations, we present our numerical results in Section~\ref{sec:numres}. The signals for deconfinement are investigated in $N_f=1$ and $N_f=2$ QCD(adj). In addition we present results for the per-site constraint effective potential of the Polyakov line phase, the adjoint Polyakov loop, and the chiral condensate in the $N_f=1$ case.

\section{Adjoint QCD on the lattice}

QCD(adj) consists of a non-Abelian gauge-field $A_\mu^c(x)$ minimally coupled to $N_f$ Majorana fermions $\lambda_i^c(x)$. The expression of the continuum action is similar to QCD
\begin{equation}
 S = \int d^4 x \left\{ -\frac{1}{4} F_{\mu\nu} F^{\mu\nu} + \frac{1}{2}\sum_{i=1}^{N_f} \bar{\lambda}_i (\slashed{D} + m)\lambda_i\right\}\, = S_g + S_f\; ,
\end{equation}
where  $F_{\mu\nu}$  is  the field strength tensor and the covariant derivative $D_{\mu}$ acts in  the adjoint representation.\footnote{A summation of colour indices is assumed.} We consider SU($N_c$) gauge groups and we only focus our numerical simulations on $N_c=2$.
Two Majorana spinors can be combined to a single Dirac spinor. Hence the counting of Majorana flavors $N_f$ corresponds formally up to a factor $1/2$ to the Dirac flavour counting used in QCD. The special case of $N_f=1$ QCD(adj) is $\mathcal{N} = 1$ supersymmetric Yang-Mills theory (SYM) and the fermion $\lambda$ is called gluino, the superpartner of the gluon. Supersymmetry is obtained in the massless limit since a finite $m$ breaks supersymmetry. 

The lattice discretization of QCD(adj) can be chosen in several different ways. We use for the discretization of $S_g$ the tree level Symanzik improved gauge action, composed of square and rectangular Wilson loops of size $1\times 2$ ($W^{(2,1)}_{\mu\nu}$) and $1\times 1$ ($W^{(1,1)}_{\mu\nu}$),
\begin{equation}
 S_g= -\frac{\beta}{N_c}\left(\frac{5}{3} \sum_{x,\mu> \nu} \textrm{Tr}\left\{W^{(1,1)}_{\mu\nu}(x)\right\} - \frac{1}{12} \sum_{x,\mu> \nu}\textrm{Tr}\left\{ W^{(2,1)}_{\mu\nu}(x) + W^{(1,2)}_{\nu\mu}(x)\right\}\right)\,.
\end{equation}

The discretization of $S_f$ is more involved. According to the Nielsen-Ninomiya theorem the implementation of a local Dirac operator on the lattice either leads to additional fermion degrees of freedom (doublers) or requires the breaking of chiral symmetry. In our approach we use a Wilson-Dirac operator which introduces an additional spin-diagonal term to decouple the doubler  modes at the cost of an explicit chiral symmetry breaking, see e.g. for a description of doublers and Wilson term \cite{Montvay:1994cy}.\footnote{
Overlap and domain wall fermions would be an alternative with a more controlled chiral symmetry breaking but their computational cost is quite demanding. Staggered fermions have been used in earlier investigations of QCD(adj), but they introduce additional degrees of freedom (tastes) and represent effectively theories with larger $N_f$ if rooting is not included.} The fermion part of our lattice action is 
\begin{equation}
S_f =
\frac{1}{2} \sum_{xy} \bar{\lambda}_x (D_w)_{xy} \lambda_y\,,
\end{equation}
with the Wilson-Dirac operator
\begin{eqnarray}
(D_w)_{x,a,\alpha;y,b,\beta}
    &=&\delta_{xy} \delta_{a,b} \delta_{\alpha,\beta}
    -\kappa \sum_{\mu=1}^{4}
      \left[ (1 - \gamma_\mu)_{\alpha,\beta}(V_\mu(x))_{ab}
                          \delta_{x+\mu,y} \right.\nonumber\\
    &&\left. + (1+\gamma_\mu)_{\alpha,\beta} (V^\dag_\mu(x-\mu))_{ab}
                          \delta_{x-\mu,y}\right]-\frac{\kappa c_{sw}}{2}\, \delta_{xy}\sigma_{\mu\nu} F^{\mu\nu}\, ,
\end{eqnarray}
where $V_{\mu}(x)$ are the link variables in the adjoint representation and $F_{\mu\nu}$ is the clover plaquette. The
hopping parameter $\kappa$ is related to the bare gluino mass $m_0$ via $\kappa=1/(2m_0+8)$. 

We have explored different strategies for the tuning of the gauge coupling $\beta = \frac{2N_c}{g^2}$ and of the parameters of the Wilson-Dirac operator. The first natural choice is the same setting as employed in earlier investigations of the bound spectrum of the theory \cite{Bergner:2015adz}. In these simulations the clover coefficient $c_{sw}$ has been set to zero and one level of stout smearing has been applied on the links $V_\mu(x)$ of the Dirac operator. In this framework $\beta=1.75$ is good compromise between the simulation cost and the control of lattice supersymmetry breaking \cite{Bergner:2013nwa,Bergner:2015adz}. In an alternative setup, we have also simulated the theory using unsmeared links and $c_{sw}=1$, such that the leading order chiral symmetry breaking effects are removed from on-shell quantities \cite{Munster:2014cja}. Different lattice actions provide the same information about the phase structure of the theory up to lattice discretization errors, enabling us to check the reliability and the consistency of our simulations. 

The integration of each Majorana fermion field yields the Pfaffian $\text{Pf}(CD_w)$, where $C$ is the charge conjugation matrix. The Pfaffian  is real but not necessarily positive on lattice. In continuum theory, the Pfaffian is real positive. 
 A sign problem appears in the Wilson formulation for odd $N_f$ in regions where the mass of the "adjoint pion",  the lightest gluino-ball, is small and lattice artefacts are dominant.  The sign problem  vanishes in the continuum limit. Based on our previous experience of lattice simulations of $\mathcal{N} = 1$ SYM, we already know in the bare parameter space where the contribution from negative Pfaffian configurations is negligible.    A positive Pfaffian is assumed in the following sections.

In numerical simulations the lattice extend is finite in all directions. Periodic boundary conditions are always applied in the three spacial directions. Our aim is the investigation of the compactified theory on a torus. Hence the extend in the spacial directions is assumed to be large enough to emulate  the infinite volume limit and the temporal direction is compactfied: 
\begin{align}
\underbrace{T^3}_{\rm large}  \times  S^1  \approx  {\mathbb R}^3 \times S^1
\end{align}
If  anti-periodic boundary conditions are  used for fermions on $S^1$ direction,  the theory on  $T^3 \times S^1$  torus emulates  the  thermal partition function on ${\mathbb R}^3 \times S^1$.    If  periodic boundary conditions are  used for fermions on $S^1$ direction,  the theory on torus corresponds to   twisted partition function \eqref{tpf} with no thermal interpretation. The latter setup can be used to realize the notion of  adiabatic continuity.

Lattice simulations have to challenge discretization and finite volume effects. Finally, numerical instabilities of the Rational Hybrid Monte Carlo algorithm forbid simulations at very small gluino masses. Hence SYM can be simulated only with non-vanishing soft supersymmetry breaking mass term. 
The continuum limit and the massless limit must be extrapolated from the numerical data.

\section{Order parameters for the phase diagram of adjoint QCD}
The lattice action is invariant under center symmetry transformations, corresponding to the multiplication of the gauge links in time direction  on a given time-slice 
with a discrete phase rotation $\exp{(i\phi_n)}$ with $\phi_n = 2\pi n / N_c$. The deconfinement transition is identified with the spontaneous breaking of this symmetry at high temperatures. The quark fields in QCD, which are fermions in the fundamental representation, break center symmetry explicitly if one imposes periodic or anti-periodic boundary conditions.  The symmetry is, however, preserved for fermions in the adjoint representation.

The Polyakov loop is the order parameter of the deconfinement phase transition. It is the path ordered product of the links in the fundamental representation along a line which wraps in the compact direction 
\begin{equation}
 P_L =\frac{1}{N_c V_3} \sum_{\vec{x}} \textrm{Tr} W^{(N_t)}(\vec{x})= \frac{1}{N_c V_3} \sum_{\vec{x}} \textrm{Tr}\left\{\prod_{t=1}^{N_t} U_4(\vec{x},t) \right\}\, ,
\end{equation}
where $V_3$ denotes the three-dimensional lattice volume of the non-compactified directions. 
The confined and deconfined phases are distinguished by a zero or non-zero expectation value of $P_L$. We consider in addition the Polyakov loop in adjoint representation $P_L^A$, the path ordered product of links in the adjoint representation. The adjoint Polyakov loop is not an order parameter of the deconfinement transition due to the screening of the adjoint color charge. More generally arbitrary windings $n$ of the Polyakov loop can be considered
\begin{equation}
 P^{(n)}_L =\frac{1}{N_c V_3} \sum_{\vec{x}} \textrm{Tr}\left[ (W^{(N_t)}(\vec{x}))^n \right]\, ,
\end{equation}
which is equivalent to consider loops in all representations. The eigenvalues of the Polyakov line $W^{(N_t)}$ play an important role in the discussions of the perturbative effective action in Section~\ref{sec:veff}.

In the massless limit,  classical SYM has an additional $U(1)$ chiral symmetry\linebreak $\lambda \rightarrow \exp{\{-\Ii \theta \gamma_5\}} \lambda$. This is not a symmetry in quantum theory due to global ABJ  anomaly. The genuine symmetry of the quantum theory is $\mathbb Z_{2N_c}$ in case of the gauge group SU($N_c$).  The  discrete symmetry is broken spontaneously to $\mathbb Z_2$, and the theory has $N_c$ isolated discrete vacua.   
The chiral condensate is the order parameter for this transition. It is defined as the derivative of the logarithm of the partition function with respect to the fermion mass, equal to the expectation value of the quark bilinear $\langle \bar{\lambda}\lambda\rangle$. This operator is multiplicatively and additively renormalized in the Wilson formulation of lattice fermions. In case of $N_f>1$ more general condensates like $\langle \det(\lambda_i \lambda_j)\rangle$ are possible \cite{Unsal:2007jx} which we have not considered in our present work. 

A susceptibility can be defined for the order parameters $O$ as
\begin{equation}\label{eq:suscept}
 \chi_O = V (\langle O^2 \rangle - \langle O \rangle^2)\,,
\end{equation}
where $V$ is the three-dimensional volume for the Polyakov loop or the four-dimensional volume for the chiral condensate. The phase transition can be identified by the peak of the susceptibility and the divergence of this peak in the infinite volume limit.
The susceptibility is subject to rather large finite volume corrections. The Binder cumulant,
\begin{equation}\label{eq:binder}
 B_4 (O) = 1-\frac{1}{3} \frac{\langle O^4\rangle}{\langle O^2\rangle^2}\;,
\end{equation}
provides a signal with smaller finite size corrections. The deconfinement transition is expected to be in the three-dimensional Ising universality class \cite{Bergner:2014saa}, and the $B_4$ value at the transition should be $0.46548(5)$ \cite{Ferrenberg:2018zst}, while in the confined and deconfined phase it is equal to zero and $2/3$ respectively.

In previous numerical simulations we have determined the deconfinement and chiral transitions in SYM at finite temperature. With thermal boundary conditions, the temperature corresponds the the inverse of the compactification radius $T=1/R$.
We have been able to observe a significantly different behaviour in the case of periodic boundary conditions for fermion fields in \cite{Bergner:2014dua}, by exploring the phase diagram in the space of the bare lattice parameters $\beta$ and $\kappa$ at fixed $N_t$. The deconfinement transition line does not intersect the critical line corresponding to the zero renormalized gluino mass. The clear separation of the two lines is already a signal of continuity, but the control of lattice artefacts and the interpretation of the phase diagram in terms of renormalized physical quantities is difficult. We follow a different approach in the present work. We keep fixed bare parameters, therefore the lattice spacing $a$ and the gluino mass are constant, and the length of the compactification radius is changed by changing the number of lattice points $N_t=R/a$ in the temporal (compactified) direction.

\section{Perturbative effective potential for the Polyakov loop on the lattice}\label{sec:veff}

The confined and deconfined phases as a function of $R$, $m$ and $N_f$ can be inferred from the effective potential of the Polyakov line, that can be calculated straightforwardly in the one-loop approximation. The effective potential is usually identified with the free energy density at a constant background gauge field $G_\mu$, which is at the one loop level determined from the quadratic fluctuations. The background is chosen such that the link in time direction $G_4$ is diagonal and all other links are set to the identity. In the case of SU(2) we have
\begin{eqnarray}
G_4 =
 \begin{pmatrix}
  e^{i\phi/N_t} & 0 \\
  0 & e^{-i\phi/N_t}
 \end{pmatrix}
\, ,
\end{eqnarray}
and Polyakov loop then reads
\begin{equation}
 P_L = \frac{1}{2}\textrm{Tr}\left\{\prod_{t=1}^{N_t} G_4 \right\} =   \frac{1}{2}\textrm{Tr}   \begin{pmatrix}
  e^{i\phi } & 0 \\
  0 & e^{-i\phi}
 \end{pmatrix} = 
  \cos(\phi)\,.
\end{equation}
Confinement (center-stability) is  realized if the minima of the effective potential for the Polyakov line phase (holonomy) $\phi$ is located at $\phi = \pi/2 \mod \pi$.
In the deconfined phase there is a minimum at $\phi=0 \mod \pi$.
For a general SU($N_c$) gauge group, the background configuration is 
\begin{eqnarray}
G_4 =\textrm{diag}\left[ e^{i\phi_1/N_t}, e^{i\phi_2/N_t},\ldots, e^{i\phi_{N_c}/N_t} \right ]\, ,
\end{eqnarray}
where $\phi_{N_c}=-\sum_{i=1}^{N_c-1} \phi_i$ and 
\begin{align}
P_L = \frac{1}{N_c}\textrm{Tr}\left\{\prod_{t=1}^{N_t} G_4 \right\} =   \frac{1}{N_c}\textrm{Tr}   \begin{pmatrix}
  e^{i\phi_1 } &  &   &\\
   & e^{i\phi_2} & & \\ 
  && \ddots   &  \\
    &&    &   e^{i\phi_{N_c}}
 \end{pmatrix}  
\end{align}

The effective potential of QCD(adj) in the one loop approximation is the sum of gluons, ghost, and fermion contributions. On the lattice the gauge links with the fluctuation field $A_\mu$ around the constant background $G_\mu$ are represented as
\begin{equation}
U_\mu(x) = G_\mu \exp\{i g A_\mu(x)\} \,,
\end{equation}
and the gluon action is expanded in a power series of $g$. 
The kinetic part of the gluon Lagrangian is 
\begin{equation}
 \mathcal{L}_{gl} = \frac{1}{2} \textrm{Tr}\left\{D^+_\mu A_\nu(x) D^+_\mu A_\nu(x)\right\}
\end{equation}
where the gauge fixing term
\begin{equation}
 S_{gf} = \sum_{x,\mu\nu} \textrm{Tr}\{ D^-_\mu A_\mu(x) D^-_\nu A_\nu(x)\}
\end{equation}
has been added to the action to fix the gauge of the field $A_\mu(x)$ while preserving gauge invariance with respect to background field $G_\mu$. Only the standard Wilson plaquette action part is relevant in this computation. $D^+$ and $D^-$ denote the forward and backward lattice covariant derivatives
\begin{eqnarray}
 D^+_\mu & = & \frac{1}{a} \left(G_\mu^\dag(x) A_\mu(x+\mu) G_\mu(x) - A_\mu(x)\right) \\
 D^-_\mu & = & \frac{1}{a} \left(A_\mu(x) - G_\mu^\dag(x-\mu) A_\mu(x-\mu) G_\mu(x-\mu) \right)\,.
\end{eqnarray}
The kinetic part of the ghost field $\eta$ is similarly
\begin{equation}
 \mathcal{L}_{gh} = \frac{1}{2} \textrm{Tr}\left\{D^+_\mu \eta(x) D^+_\mu \eta(x)\right\}\,;
\end{equation}
the last contribution comes from the Wilson fermion action ($m_0=am$)
\begin{equation}
 \mathcal{L}_{f} = \frac{1}{2}\sum_{f = 1}^{N_f} \bar\lambda^f(x) \left\{\gamma_\mu \frac{D^+_\mu + D^-_\mu}{2} + a r D^+_\mu D^-_\mu + m\right\} \lambda^f(x)\,.
\end{equation}

The action is quadratic in fluctuating fields $A_\mu$, $\eta$ and $\lambda$ in the one-loop approximation, therefore these fields can be integrated out leading to
\begin{align}
V_3 V_{\textrm{eff}} &= \left(\frac{4}{2} - 1 \right) \log \det(D_\mu^- D_\mu^+) - \frac{N_f}{2} \log \det\left(\gamma_\mu \frac{D^+_\mu + D^-_\mu}{2} + a r D^+_\mu D^-_\mu + m\right)\nonumber \\
&= \log \det(D_\mu^- D_\mu^+) - N_f \log \det\left(\frac{1}{4}(D^+_\mu + D^-_\mu)^2 + (a r D^+_\mu D^-_\mu + m)^2\right)\,.\label{vefflattice}
\end{align}
Each boson field contributes to $V_{\textrm{eff}}$ with a prefactor $+\frac{1}{2}$, and each fermion field with a prefactor $-1$. The first term comes from the gauge part of the action, with a prefactor which counts four boson fields $A_\mu$ minus the fermion ghost field $\eta$. The last term is the contribution from the Majorana fermions with an additional factor $\frac{1}{2}$ due to the fact that integrating our fermion induce a  Pfaffian.

The effective potential \eqref{vefflattice} shows clearly how a mismatch between fermion and boson contributions is introduced by the lattice discretization. The Wilson fermions have a different derivative operator and an additional momentum dependent mass term in order to remove the doubling modes.  
In the continuum, a more compact expression for $V_{\textrm{eff}}$ is obtained in the massless case. It can be recovered in the naive $a \rightarrow 0$ limit of \eqref{vefflattice}
\begin{equation}
 V_3 V_{\textrm{eff}} =  (1-N_f) \log \det(D^2)\, ,
\end{equation}
since $\log\det(\slashed{D}) =  \frac{1}{2} \log\det(\slashed{D}\slashed{D}) = \frac{1}{2} \log \det(D^2)$. In this limit the difference between the boson and fermion derivative operators disappears.
Gluon and adjoint fermion fields have an opposite contribution to the one-loop effective potential of the Polyakov loop which cancel exactly in the continuum if $N_f = 1$, i.~e.\ in the case of the $\mathcal{N}=1$ SYM.  
Supersymmetry 
 ensures that this result holds to all orders of perturbation theory.  In continuum  $\mathcal{N}=1$ SYM theory, center stability  is driven by pure non-perturbative effects  \cite{Poppitz:2012sw}. For $N_f>1$ continuum theory, the center-stability is a one-loop perturbative effect. However,  
 the symmetry between gluons and gluinos in SYM is violated on the lattice,  and $V_{\textrm{eff}}$ differs from zero for non-zero lattice spacings. This fact will play an interesting role in  Section \ref{explanation} in order to explain the difference of  lattice and continuum phase diagram.

The effective potential of the Polyakov line holonomies $\phi_a$ (or their differences $\phi_{ab}=\phi_a-\phi_b$) can be further simplified in momentum space \cite{Bringoltz:2009mi, Poppitz:2009fm}.
Assuming an infinite spacial extend of the lattice, an $\infty^3 \times N_t$ site lattice, and neglecting  holonomy independent  contributions, we obtain the following representation
\begin{align}
V_{\textrm{eff}}(\{\phi_a\}) &=\sum_{a\neq b} V_{\textrm{eff}}(\phi_{ab})= \sum_{k= 0}^{N_t-1 }\sum_{a\neq b}  \, \int_{-\pi}^\pi \left(\frac{dp}{2\pi}\right)^3\, \left\{\log \left[  \hat p^2 + 4\sin^2\left(\frac{\phi_{ab} + 2 \pi k  }{2N_t}\right)  \right]\right.\nonumber \\
&\left. - N_f \log\left[    {\hat{\hat{p}}}^2 + \sin^2 \left( \frac{ \phi_{ab} + 2\pi k  }{N_t} \right) +  
\left( m_0 + \frac{r}{2}  \left[ \hat p^2 + 4 \sin^2\left(\frac{\phi_{ab} + 2 \pi k  }{2N_t}\right)    \right]   \right)^2
 \right]\right\}. \qquad \qquad 
\label{1loop}
\end{align}
where  the lattice momenta $\hat p$ and $\hat{\hat{ p}}$ are defined as
\begin{align}
\hat p^2 = \sum_{i=1}^3  4 \sin^2 \left(\frac{p_i }{ 2} \right)\quad \text{and}\quad  \hat{\hat{p}}^2 = \sum_{i=1}^3 \sin^2 p_i\, .
\end{align}
Consider first $r=0$ and $m_0=0$, and set the holonomy  field to zero in  \eqref{1loop}.
The difference between boson and fermion derivatives on the lattice is essentially  reflected in  $\hat p^2$ vs. $\hat{\hat{ p}}^2$.   In fact, for bosons, the inverse propagator   only vanishes at  the  origin of the   Brillouin zone, as in the continuum case. 
However, for fermions, the inverse propagator is zero at all 16 corners of Brillouin zone, $(0, 0, 0, 0), (\pm\pi, 0, 0, 0), \ldots , 
 (\pm\pi, \pm\pi, \pm\pi, \pm\pi )$. The corners  except the origin are referred to as doublers, and undesired from the continuum point of view. 
On the lattice, the Wilson term proportional to $r$ leads to a momentum dependent mass that removes the doublers in the continuum limit.  The Wilson term lifts the doublers by making the modes at $p_{
\mu} \sim \pi$ acquire a mass of the order of cut-off (inverse lattice spacing). However, the Wilson term explicitly breaks chiral symmetry, and the fermion mass is both additively and multiplicatively renormalized.

 The difference to the fermion determinants of the free theory is a shift of the modes in the compact or temporal direction by $\phi_{ab}$. 
This generalizes to any kind of fermion operator that can be implemented on the lattice: the effective potential is easily derived from a shifted momentum space representation. The shifts for the adjoint representation are $\frac{\phi_{ab}}{N_t}$, for the fundamental one they would be $\frac{\phi_a}{N_t}$.
In the case of SU(2) there is only one shift for the adjoint representation: $\phi_{12}=-\phi_{21}=2\phi$. 

The effective potential for $\phi_{ab}$ can be translated into an effective potential of different windings of the Polyakov loop. Again up to constant expressions one obtains the quadratic potential
\begin{align}
V_{\textrm{eff}}(\{\phi_a\}) = \sum_{n=1}^{\infty}  \sum_{a\neq b}\VR\; e^{ir\phi_{ab}} =N_c^2\sum_{n=1}^{\infty}  \VR |P_L^{(n)}|^2 + \text{const}\; .
\label{1plr}
\end{align}
The coefficients $\VR$ are the Fourier transform of $V_{\textrm{eff}}(\phi_{ab})$ with respect to $\phi_{ab}$
\begin{align}
 \VR =  \int_0^{2\pi} \frac{d \phi_{ab}}{2\pi}\;  e^{-i r \phi_{ab}}\;  V_{\textrm{eff}}(\phi_{ab})\, .
\end{align}
After some algebra, the following expression is obtained for the coefficients $\VR$
\begin{multline}
\VR = N_t \int_{-\pi}^\pi \frac{d\omega}{2\pi}\,e^{ir N_t \omega}\, \int_{-\pi}^\pi \left(\frac{dp}{2\pi}\right)^3\\  \log   \left( \frac{  \left[  \hat p^2 + 4\sin^2\left(\frac{\omega}{2}\right)\right] }
{ \left[   \hat{\hat{p}}^2 + \sin^2\left(\omega\right) +  \left(  m_0 +  \frac{r}{2}\left( \hat p^2 +  4  \sin^2\left(\frac{\omega}{2} \right)\right)   \right)^2  
\right]^{N_f} } \right)
\label{1lpvr}
\end{multline}
If the $\VR$ are positive for $n \leq \lfloor {N_c \over 2}  \rfloor$, $\mathbb Z_{N_c}$ center-symmetry is preserved and 
the minimum of the potential is at    $P_L^{(n)} =0$  for  $n \neq 0 \mod N_c $, i.e,  the expectation value of the Polyakov loops 
with winding number $n \neq 0 \mod N_c $ 
is zero. 
 
If some of the    $\VR$ are negative for $n \leq \lfloor {N_c \over 2}  \rfloor$,  a corresponding subset of Polyakov loops 
will develop non-zero expectation values. In fact, if  $\VR<0$ for  $n=1$, then so are all higher $n$.  In this case,  $\mathbb Z_{N_c}$ center-symmetry is completely broken.  If  $\VR<0$ for  $1 \leq n \leq k \leq  \lfloor {N_c \over 2}  \rfloor $, then the center symmetry may break to different discrete subgroups that can be easily determined. 

Note that an increase of $\frac{1}{\VR}$ indicates also an increasing susceptibility of the Polyakov line with winding number $n$. This provides a signal for the effective potential beyond the perturbative level that is investigated in Section~\ref{sec:numres}.

\begin{figure}
\centering
\includegraphics[width=.57\textwidth]{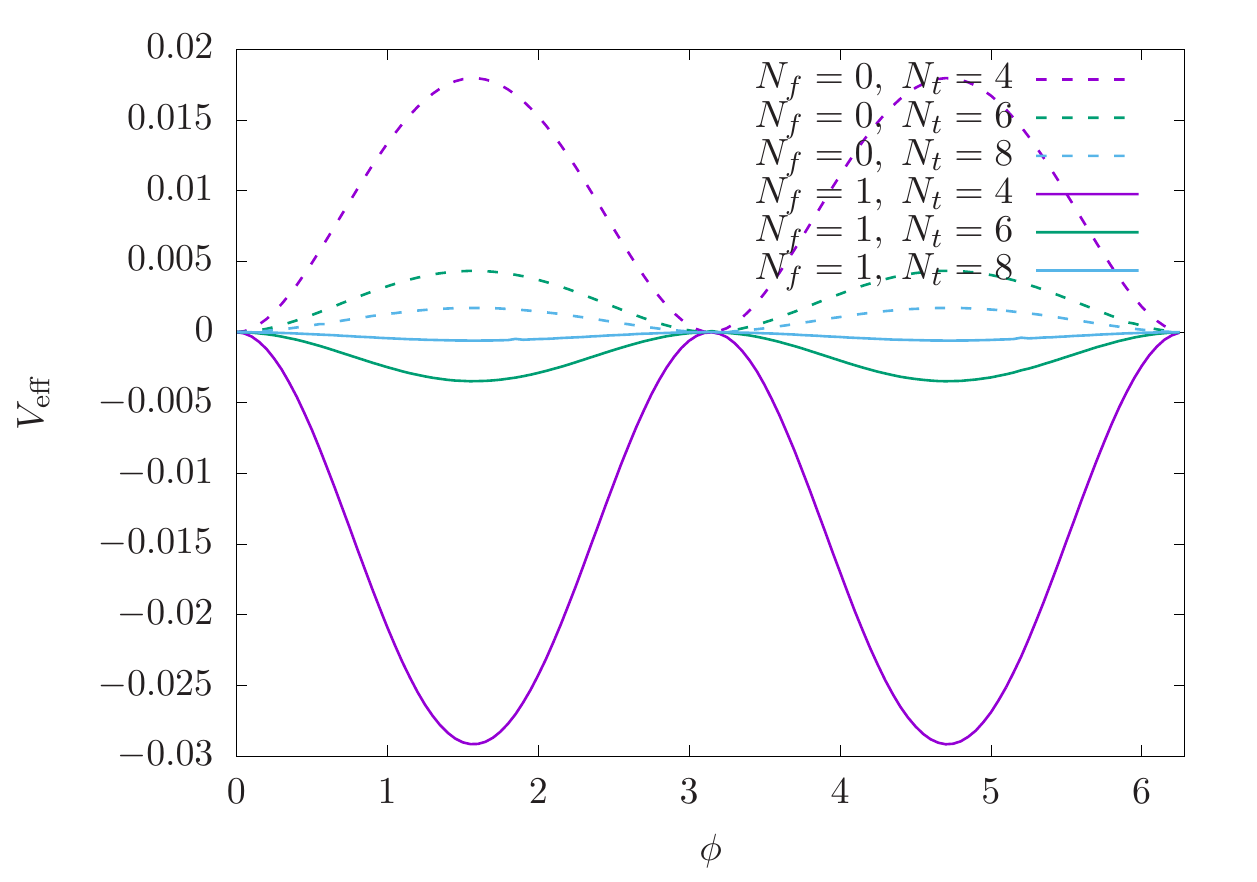}
\caption{One loop effective potential for the Polyakov loop phase $\phi$ in SU(2) SYM on the lattice for different $N_t$ at $m_0=0$ comparing pure gauge ($N_f=0$) and SYM ($N_f=1$).  The SYM effective potential is different from zero due to lattice artefacts. It is confining and the expected flat behavior is reached only asymptotically at large $N_t$. The potentials are normalized to zero at $\phi=0$ in this plot.}\label{one-loop-veff}
\end{figure}
We have computed the one-loop effective potential of SU(2) SYM on the lattice \eqref{vefflattice} or equivalently \eqref{1loop} for several values of $N_t$.  
Clearly, unlike the continuum supersymmetric theory, for which due to supersymmetry the effective potential is zero to all orders in the perturbative  expansion, the one-loop potential in the lattice formulation \eqref{vefflattice}, \eqref{1loop}  is actually non-vanishing,  
\begin{align}
& V_{\textrm{eff, continuum}}(\{\phi_a\})  =0,      \cr
& V_{\textrm{eff, lattice}}(\{\phi_a\}) =   N_c^2\sum_{n=1}^{\infty} \VR |P_L^{(n)}|^2\; {\rm with}\; \VR>0  \; {\rm for} \;  n \leq \Big\lfloor {N_c \over 2}  \Big\rfloor\;.
\end{align}
This is hardly surprising because the lattice formulation based on Wilson fermions does not respect supersymmetry, and as such, we indeed expect a potential to be induced at  one-loop  order. The correct continuum limit is approached as $R/a$ goes to infinity, which means the lattice spacing goes to zero at fixed physical $R$.

Remarkably,  Wilson fermions with Wilson parameter $r=1$, even when the effect of doublers is completely lifted,   have a stronger confining effect than continuum fermions, see Figure~\ref{one-loop-veff}, and lead to center stability even at one-loop level. In this respect, 
lattice SYM is similar to the  $ N_f>1 $ continuum QCD(adj) \cite{Unsal:2010qh}.   The  one-loop  prediction is a center-symmetric (confined)  phase in the small $R$ regime at a given fixed lattice spacing and $m_0=0$. 
In the small $R/a$ regime, the dynamics probes higher scales towards the cutoff $1/a$. At these scales the doubling modes that are lifted with masses of the order of the cutoff are still relevant and have a confining effect.

If the mass is increased, a phase transition to a deconfined phase  occurs \cite{Unsal:2010qh}.   This is calculable both in lattice perturbation theory and continuum methods and is discussed in Section~\ref{phase}. 
This phase is connected to the deconfined phase in pure Yang-Mills theory. At $N_c>2$ the deconfinement transition is replaced by a transition to several intermediate phases with partially broken gauge symmetry, corresponding to the Higgs phases in the Hosotani mechanism \cite{Cossu:2013ora}. The deconfined phase connected to the pure Yang-Mills case is reached after crossing these additional phases when increasing $R$.

\vspace{3mm}
\noindent 
\subsection{Abelian vs. non-Abelian confinement regimes: first pass} 
In the following we discuss in more detail the particular features of confinement mechanism at small radius that arise in QCD(adj) for $N_f>1$ and at a finite lattice spacing also in the $N_f=1$ case.

The unbroken center symmetry at any $R \in [0, \infty)$ does not have a {\it unique}  implication for Polyakov loop eigenvalues.   Originally, Polyakov thought that the phases of the loop would have completely random fluctuations and this would be the  origin of the a vanishing expectation value for the Polyakov line operator in the confined phase. However, from the analysis of the confining perturbative effective potential we obtain two different pictures. Indeed, both at small and large-$R$, $\langle P_L  \rangle =0 $, but at small-$R$ 
the (untraced) Polyakov line is essentially  $ \begin{pmatrix}
  e^{i \frac{\pi}{2}  } & 0 \\
  0 & e^{-i \frac{\pi}{2} }
 \end{pmatrix}$ with small-fluctuations around it. We refer to this regime as Abelian confinement regime, where the theory dynamically abelianized due to the induced potential \eqref{vefflattice}, \eqref{1loop}.\footnote{ \;\; Dynamical abelianization 
 is interesting on its own right, and the class of theories we are examining exhibits some differences from other calculable 
 QFTs which exhibit confinement.    In Georgi-Glashow model in 3d \cite{Georgi:1972cj,  Polyakov:1976fu, Anber:2013xfa}  and Seiberg-Witten theory in 4d \cite{Seiberg:1994rs},  the long distance theory abelianize due to  a tree-level (classical)  potential.   In QCD(adj), at tree level, there is no potential for Polyakov loop. It is induced at one-loop level dynamically, and leads to  abelianization of the long distance  theory \cite{Unsal:2007jx}. Dynamical abelianization should also not be confused with 't Hooft's   maximal abelian gauge proposal \cite{tHooft:1981bkw}, which enforces abelianization by hand as a gauge choice. Needless to say, this is done in QFT which becomes strongly coupled at large-distance and in particular, is not semi-classically calculable.} 

At large-$R$, the theory is strongly coupled at the compactification scale and the fluctuations of the Polyakov loop eigenvalues are random as can be seen from simulation results presented in Section~\ref{sec:effpot}.    
We refer to this regime as non-Abelian confinement regime.

Some intuition for the large-$R$  case  may be gained from strong coupling lattice perturbation theory. However, note that the strong coupling here refers to bare strong coupling at the cut-off. This regime of lattice theory is not necessarily continuously connected to continuum physics, yet the  strong coupling expansion has some benefits 
to understand the lattice results.  In particular, at infinite bare coupling,  $\beta=0$,  we can ignore the action and the 
 first term in the strong coupling limit is the reduced Haar measure with an effective potential for the Polyakov line eigenvalues
\begin{align}
 V_\text{Haar}=-\frac{1}{2}\sum_{a\neq b}\log\left[\sin^2\left(\frac{\phi_{ab}}{2} \right)\right]\,.
\end{align}
 The reduced Haar measure will always drive the theory to the confined phase.  Since we are discussing the continuum physics that arise from bare  weak  coupling (as per asymptotic freedom) of  lattice field theory,  we will not discuss strong coupling expansion.  Nevertheless the Haar measure has to be considered as an important contribution in the lattice simulations. In simulations,  we use  the distribution of the  per-site Polyakov loop  phase $\phi$ normalized with respect to the Haar measure for different $R$ to gain a sense of distinction between Abelian and non-Abelian confinement regimes. 

The picture that  emerges by the examination of the  thermal and graded partition functions  corresponding to thermal and periodic compactification 
with analytic (lattice perturbation theory at weak coupling domain) and numerical  (both at weak and strong coupling domain) methods is as follows: 
At low temperatures or large-circle size $R$ where the theory becomes strongly coupled,    there are rather unconstrained fluctuations of the $\phi_{ab}$, which leads to a vanishing Polyakov line expectation value.  
At higher temperatures,  or small circles, the weak coupling at the scale of the circle size admits a calculable potential for the Polyakov loop, which may lead to either to a confined or deconfined vacuum depending on the details. 
The Polyakov line expectation value is in this case determined by the minimum of the effective potential of $\phi_{ab}$ and the {\it randomness} of the  fluctuations becomes suppressed, regardless of whether the theory is confined or deconfined. However, the suppression of the randomness of fluctuations will happen in different ways: in the thermal deconfined phase, there is a dominance of the $\phi=0$ center-broken configuration, in the periodic confined phase, a dominance of the $\phi=\frac{\pi}{2}$ center-symmetric  configuration. The center-symmetric suppression of the randomness of fluctuations is a signal of the  dynamical abelianization according to the Abelian confinement picture.  The confinement that takes place 
at large-circle (strong coupling) is   referred to as  non-Abelian confinement picture. In the present work, we provide evidence for the continuity between the two. 

It is also  worth mentioning that historically better known examples of Abelian confinement mechanisms such as   3d Polyakov model \cite{Polyakov:1976fu}  and softly broken Seiberg-Witten theory in 4d \cite{Seiberg:1994rs},  have a property which is unlike pure Yang-Mills. In pure YM, $k$-string tensions are classified by $N_c$-ality.  There is only one-type of string between a quark with $N_c$-ality $k$ and  its anti-quark  with tension $T_k$. Consider for example the 1-string tension $T_1$. However, in the above mentioned theories, since the gauge group is $U(1)^{N_c-1}$ and since the Weyl symmetry  permuting gauge group factors are spontaneously broken, 
$T_1$ is replaced by $N_c-1$ fundamental strings, $T_{1,j}$ as shown in \cite{Douglas:1995nw}.    This is not so in QCD(adj) as well as deformed YM, where the $\mathbb Z_{N_c}$ sub-group of the Weyl group remains intact in the Abelian confinement regime.
Consequently, there is again only one-type of fundamental string, see  \cite{Poppitz:2017ivi} for details. 

\subsection{Different Polyakov line effective actions}\label{sec:plactions}
A better understanding of the different non-perturbative definitions of the Polyakov line effective action is required to link lattice results to the one-loop perturbative calculations. There is a difference between th effective action for the Polyakov loop $P_L$ and the one of $\phi_a$. These two effective actions are related for a constant background by the Fourier representation of Eq.~\eqref{1plr}, but there is no general identification between the expectation values $\langle P_L\rangle$ and $\langle \phi_a \rangle$.  We won't enter the discussion about the convexity of the effective action since it is not relevant for our discussions, see \cite{ORaifeartaigh:1986axd} for further details.

The relevant counterpart of the effective action that is easily accessible on the lattice is the constraint effective potential, see \cite{ORaifeartaigh:1986axd,KorthalsAltes:1993ca,Dumitru:2013xna} for a discussion. Up to an overall constant, it is the logarithm of the probability density
of the observable $O$
\begin{align}\label{ceffpot}
 V_{\text{c}}(\Phi)=-\frac{1}{V}\log\left[\int \mathcal{D}\theta\; \delta(\Phi-O(\theta))e^{-S[\theta]}  \right]=-\log\left[P_{pdf,O}(\Phi)\right]+\text{constant}\, .
\end{align}
The integration variable $\theta$ represents all gauge and fermion fields. The probability density $P_{pdf,O}(\Phi)$ can be numerically approximated by the histogram of $O$ obtained from the Monte-Carlo data. 
The most relevant information can be already extracted from the lowest moments of the distribution. A crucial role is played by the second moment which is equivalent to the susceptibility of Eq.~\eqref{eq:suscept}. The susceptibility is the inverse of the coefficient in front of the quadratic term in the effective potential. 

The definition \eqref{ceffpot} shows that the identification of the phase to the Polyakov line effective potential in Eq.~\eqref{1plr} is unambiguous only in the constant field limit. In the more general case the Polyakov line eigenvalues and the phases are space-dependent. The $\phi_a$ in the effective potential can be identified with the volume averaged phase factors or the phases of the volume averaged Wilson lines. In both cases the direct application of the perturbative formulas is not possible since either the volume averaged loop can not be determined from the volume average phases, or the volume averaged Wilson line is not a group element. A qualitative agreement with the potential of Eq.~\eqref{1plr} is still expected, especially in the strong and weak coupling limit where the fluctuations of the phase factors are suppressed.

A constraint effective potential can also be defined from the per-site distribution of the observable. This distribution can be determined with a better accuracy due to the additional volume averaging factor in the statistical sampling. The per-site distribution is significantly different from the distribution of the volume averaged observable. Smoothing techniques of the field configuration can be applied in order to match the two distributions, but still the infinite volume limit is not well defined for the per-site distribution. In the strong coupling limit the per-site constraint effective potential agrees with the constraint effective potential of the Polyakov line. This leads to the several different definitions of the Polyakov line effective potential mentioned in the literature: the Polyakov line effective potential, the per-site Polyakov line effective potential and the same for the effective potential of $\phi_a$.

The per-site distribution has been investigated in particular to confirm the non-Abelian confinement. As explained above, the contributions of the Haar measure have to be factored out in order make contact to this view of the confinement mechanism. Such an analysis has been done in \cite{Cossu:2013ora} for the case of SU(3) QCD(adj) with larger $N_f>1$ and we perform a similar analysis here for SU(2) SYM. In addition, we discuss the susceptibility in order to get information about the form of the effective action from the numerical simulations.

\section{The phase diagram on the continuum and  lattice}
\label{phase}
\begin{figure}
\begin{center}
\subfigure[SYM continuum\label{phasediag2-a}]{\includegraphics[width=.32\textwidth]{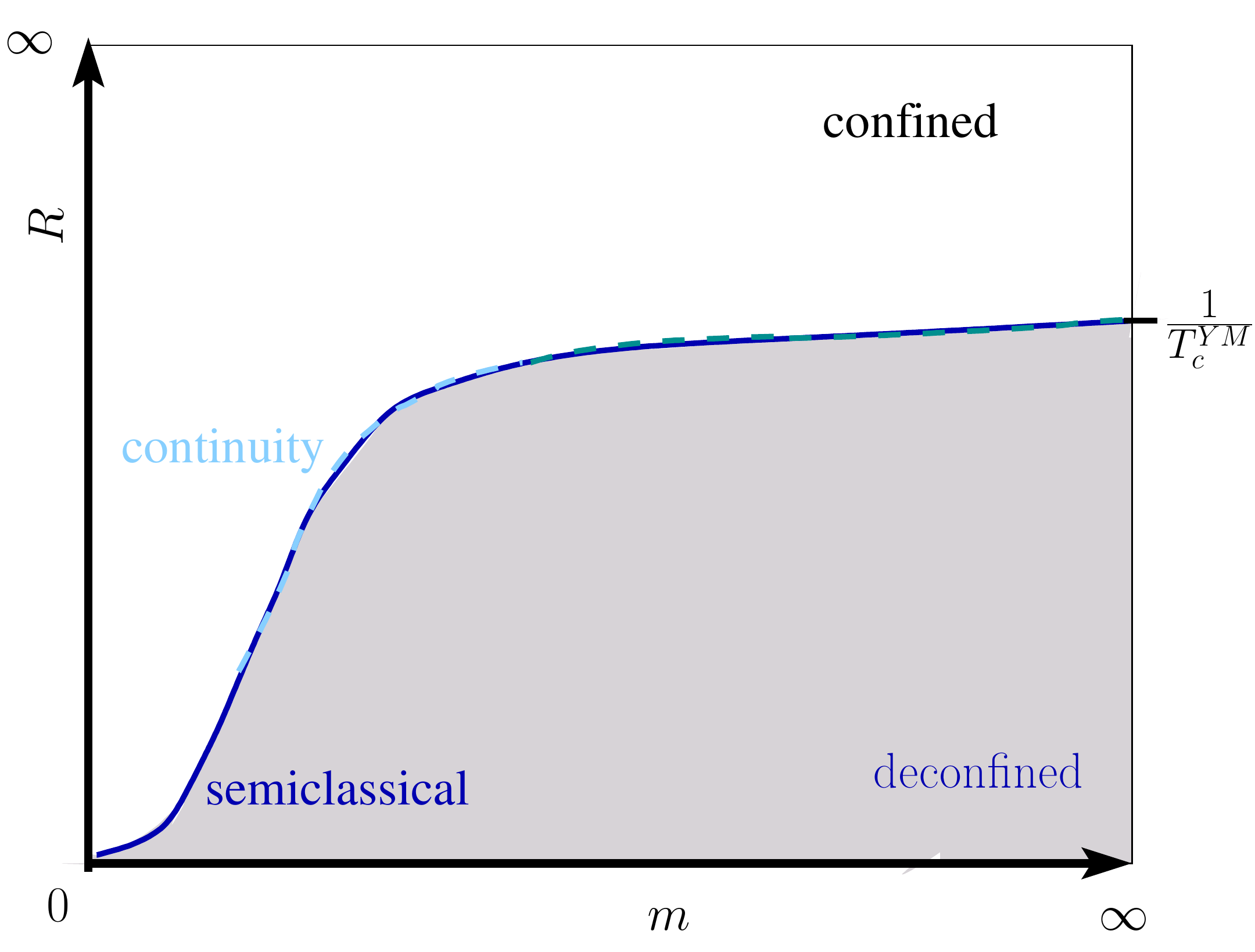}}
\subfigure[SYM lattice\label{phasediag2-b}]{\includegraphics[width=.32\textwidth]{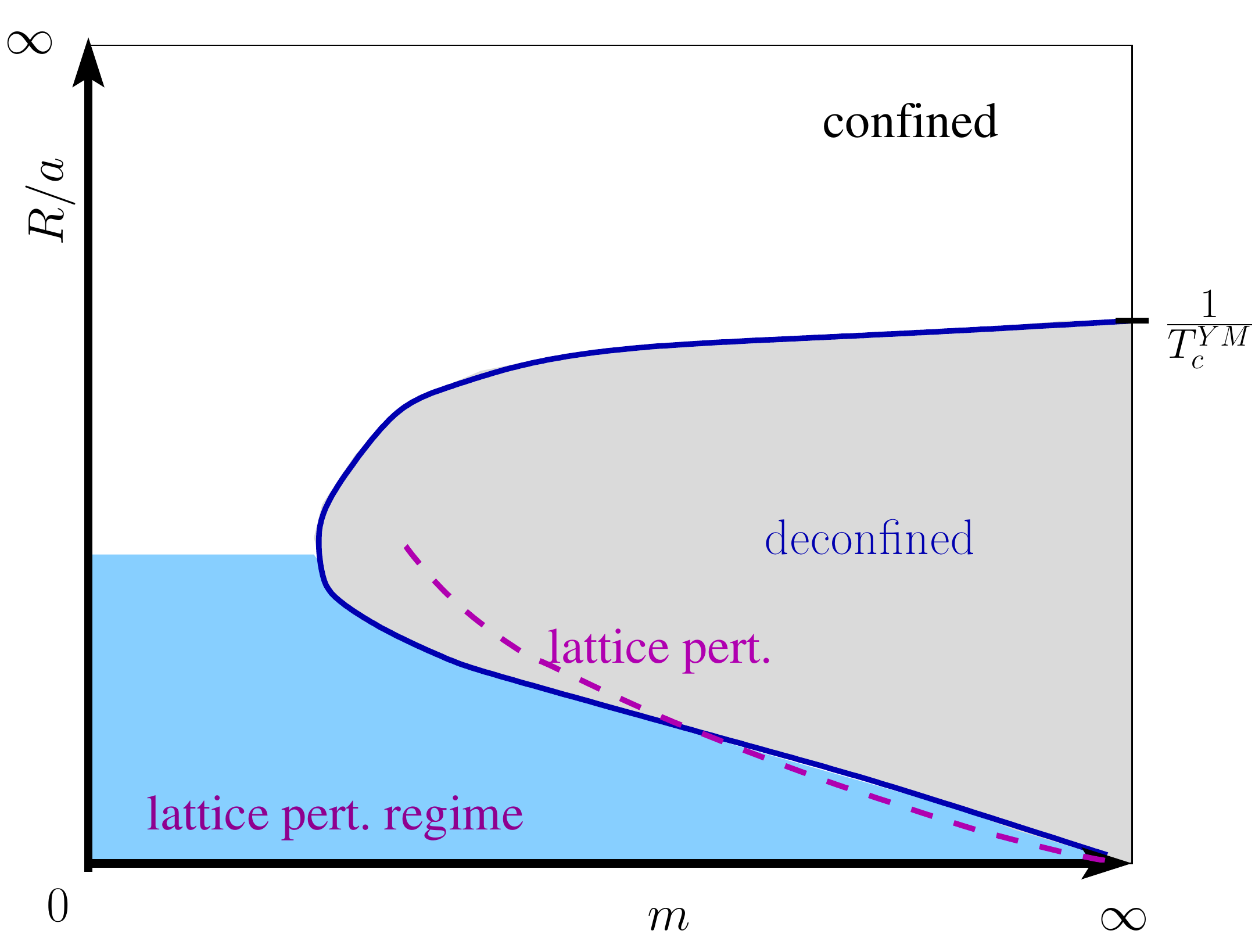}}
\subfigure[$N_f^{*}<N_f<N_f^{\rm a. f.}$\label{phasediag2-c}]{\includegraphics[width=.32\textwidth]{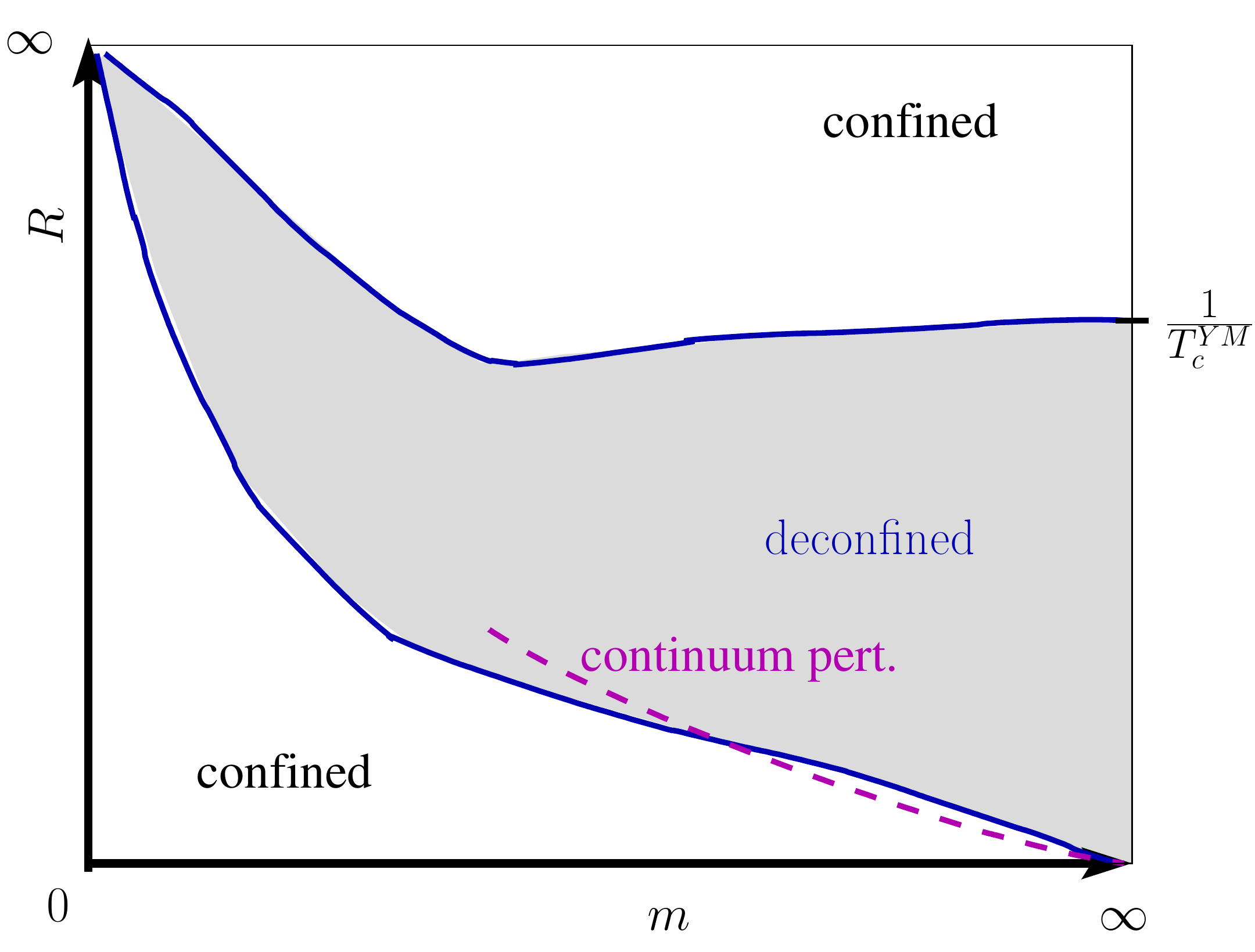}} 
\end{center}
\caption{A sketch of the expected phase diagrams in $N_f$ QCD(adj). Figure (a) is the continuum case in SYM with the semiclassical predictions valid on the small $(R, m)$ corner, the well established numerical results  at large-$m$  (where semi-classic does not apply)
and the 
conjectured continuity of the transition line.   
 Figure (b) is the phase diagram of SYM on the lattice at a fixed lattice spacing.  At small $R/a$ 
regime the transition line approaches lattice perturbation theory. The ``lattice pert.\ regime'' corresponds to a region where the 
coupling stays small at the scale of compactification.  There are crucial 
discretization effects  in this regime.    In this regime the theory behaves similar to continuum QCD(adj) with larger $N_f$.
 (Figure (c))
If $N_f^{*}<N_f<N_f^{\rm a. f.}$ becomes larger than a certain critical value  dictated by lower end of conformal window and smaller than asymptotic freedom  bound, the theory on decompactification limit is a CFT.  However, upon compactification (with periodic boundary conditions), center is stable and confinement sets in. Continuum one-loop perturbative analysis   can be used to show that $R_c m =3.484$ for $N_f=5$ (lower transition line). The upper transition line is non-perturbative and not calculable.
}\label{phasediag2}
\end{figure}

Below, we review the phase diagram of  ${\cal N}=1$ SYM and QCD(adj) with periodic boundary condition on $\mathbb R^3 \times S^1$. Part of these phase diagrams are calculable by weak coupling methods involving a combination of  perturbative one-loop effective potential for Wilson line and non-perturbative semi-classical methods.   

Next, we describe the calculable  critical compactification radius $R_c$ as a function of $N_f$ and $m_0$ by using the  one-loop effective potential on the lattice.  We also  briefly review the part  of the phase diagram  that is accessible by lattice techniques, but not via weak coupling methods. 

\vspace{3mm}
\noindent
{\bf Continuum, ${\bf  N_f=1 }$ QCD(adj): }  
The phase diagram of ${\cal N}=1$ SYM theory  in continuum is shown in  Figure~\ref{phasediag2-a}.  The small $(m, R)$ corner of this phase diagram  is semi-classically calculable  and exhibits a  center-symmetry  changing phase transition thanks to asymptotic freedom and unbroken center symmetry in the chiral limit\cite{Poppitz:2012sw,Unsal:2010qh}. 
 In the large-$m$ regime, 
 the theory approaches to pure YM as the fermion decouples.  This regime is not semi-classically calculable, but from lattice simulations, it is known that 
 a transition exists and it is natural to expect that  the transition line extrapolates from small $m$ to large-$m$ regime.   
 In this picture, there is only one transition line from confined to deconfined phase.  
  The theory is non-perturbatively confined for all $m<m_c$  or $R> R_c$ in the calculable  limit where  the transition boundary is at \cite{Poppitz:2012sw}
\begin{align} 
 R_c= \Lambda^{-1} \sqrt { m \over 8 \Lambda}   \; \; \;  {\rm  for}  \; \; \;   N_f=1
 \end{align} 
where $\Lambda$ is the strong scale.  

There are two  conjectural  continuities in the phase diagram  shown in   Figure~\ref{phasediag2-a} corresponding to the twisted partition function \eqref{tpf}:
\begin{itemize} 
\item Adiabatic continuity between the  small-$R$ weak coupling confined phase and 
large-$R$ regime where the  long distances theory becomes strongly coupled. 
\item The continuity of semiclassical (weak coupling)  and the  strong coupling  phase transition.
\end{itemize}
There is a sufficiently strong reason to believe continuity if the fermion mass is zero. \eqref{tpf} reduces down to supersymmetric Witten index \cite{Witten:1982df} at $m=0$,
\begin{align}
I_W=\tilde Z(R, m=0) =    \tr [e^{-R \hat{H}(0)} (-1)^F]  =N_c \qquad {\rm for} \; \; \mathrm{SU}(N_c),
 \end{align}
which counts the number of ground states.  At fixed small-$m$, for $R<R_c$, one can prove the stability   of the confined phase, and also prove the existence of center symmetry breaking for $R<R_c$ \cite{Poppitz:2012sw}.  
The  main aim of the present work is to find a non-perturbative evidence for the above adiabatic continuity  conjectures. 

\vspace{3mm}
\noindent
{\bf$\bm{1 < N_f <N_f^{a.f.}}$ QCD(adj):}
Let us first recall the conjectured phase diagram of QCD(adj) in the continuum.   Let  $N_f^{*}$  denote the putative number of flavors below which the theory is confining  and above which, it exhibits  IR-conformality on  ${\mathbb R}^4$. For $N_f\geq N_f^{a.f.}$ even asymptotic freedom is lost.

The expected phase diagram 
for ${  1 < N_f <N_f^{*}}$  is the same as shown in  Figure~\ref{phasediag2-b} and for ${N_f^* < N_f <N_f^{a.f.}}$, it   is    shown in  Figure~\ref{phasediag2-c}.  In both of these diagrams, upper transition lines are incalculable by weak coupling methods and lower lines are calculable. The one-loop potential is of the form:
 \begin{align} 
& V^{\rm pert.}  (\phi) = N^2 \sum_{n=1}^{\infty} V_n |P_L^{(n)}|^2, \qquad  V_n=  \frac{ 4} {\pi^2 n^4} \left[ -1+ \frac{N_f}{2} (n R m)^2 K_2(nRm) \right].
 \end{align} 
Even at  very large-$m$,  as one dials $R$, it is possible to show that 
$V_1$ is positive for $R< R_c$ and center symmetry is stable.   $V_1$ is negative for $R_c<R \lesssim  \Lambda$. This calculable phase transition correspond to the lower lines  in Figures~\ref{phasediag2-b} and ~\ref{phasediag2-c}.  
The lower phase  boundary for $\mathrm{SU}(2)$ $N_f$ flavor QCD(adj) for any  $N_f < N_f^{a.f}$    is given by\cite{Unsal:2010qh}
\begin{align} \label{eq:contmc}
 m_c  R_c^{\rm cont.} = \{2.027, 2.709, 3.154, 3.484 \}  \; \; \; {\rm  for}\;  \;\;   N_f=2,3, 4, 5.
 \end{align} 
 
The interesting aspect of the phase diagram is  the { \it non-decoupling } of heavy fermions once the circle  size is  taken sufficiently small. In that regime, the fermion, despite being heavy, enters in the combination $mR$ and a sufficiently small $R$ makes the fermions behave as if they were light in that regime.    The upper lines in  these figures are  incalculable by weak coupling methods, but it is essentially the deconfinement radius of pure YM theory because massive fermions decouple in that regime.

\begin{figure}
\begin{center}
\subfigure[Phase transition\label{latphasetrans1}]{\includegraphics[width=.4\textwidth]{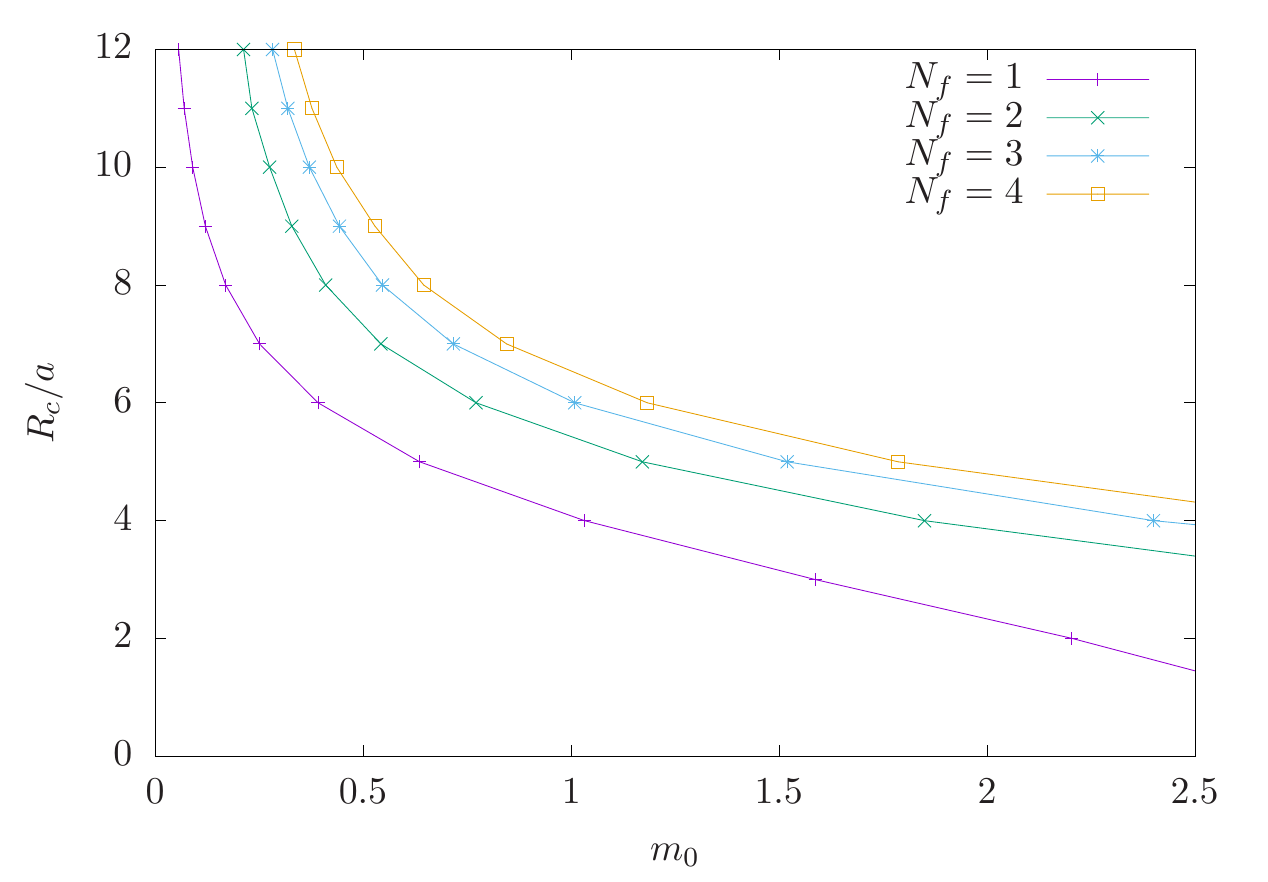}}
\subfigure[Continuum limit\label{oneloopcontlim}]{\includegraphics[width=.4\textwidth]{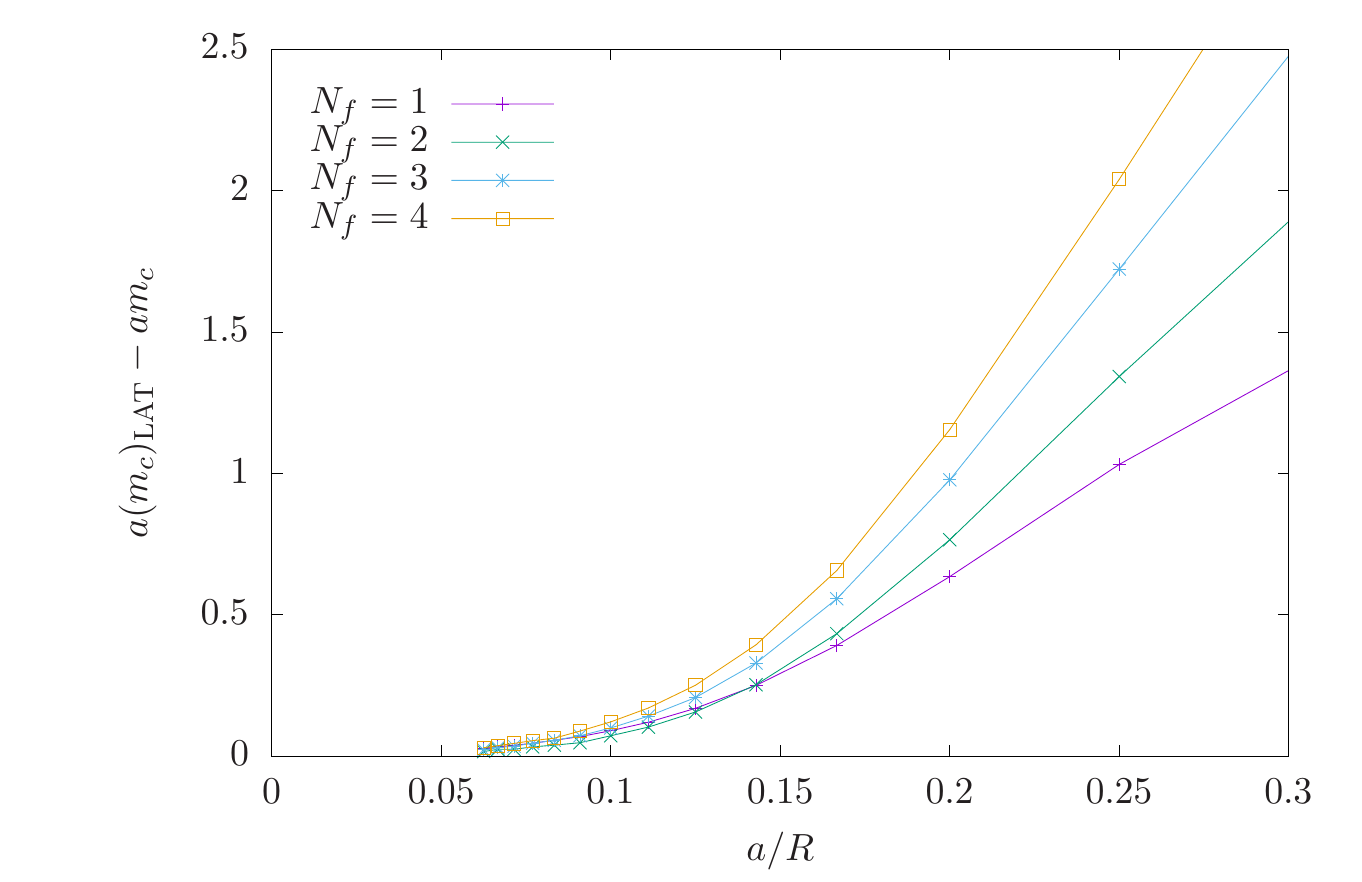}}
\subfigure[Transition as function of $R/a$\label{fixednt}]{\includegraphics[width=.4\textwidth]{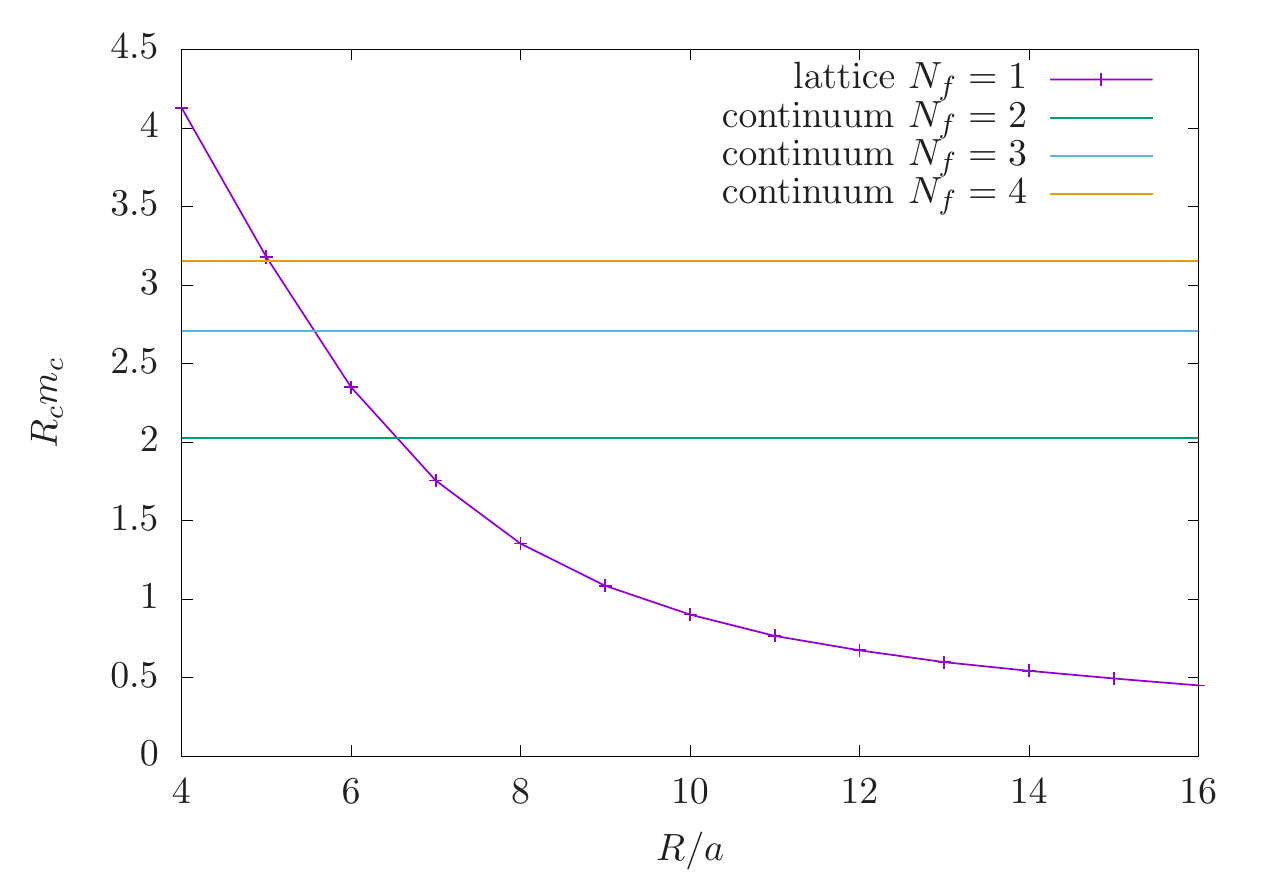}}
\end{center}
\caption{Figure (a): Numerical determination of the critical radius $R_c$ for the deconfinement transition as a function of the mass in the one-loop approximation on the lattice from Eq.~\eqref{1loop} for $N_f$ QCD(adj). 
In Figure (b) the deviations of the critical mass $(m_c)_{\mathrm{LAT}}$ determined by the one-loop lattice effective potential from the expected continuum result $m_c$ (Eq.~\eqref{eq:contmc}) are represented as a function of $a/R$. Keeping $R$ fixed, the difference tends to zero in the continuum limit. Figure (c) represents the transition for $N_f=1$ as a function of $R/a$ to compare $R_c m_c$ with the values of Eq.~\eqref{eq:contmc}.}
\label{phasediag1}
\end{figure}

\vspace{3mm}
\noindent
{\bf Lattice:}
The expected deconfined regions on the lattice are shown for different $N_f$ in Figure~\ref{latphasetrans1}. Even for $N_f=1$, there is a perturbative transition line since lattice artefacts violate the exact cancellation between gluon and fermion loops, especially at scales close to the lattice cutoff $1/a$.  Such a significant deviation from the continuum phase diagram is expected for small $R/a$, while in the large $R/a$ regime a good agreement with the continuum predictions should be observed. The corresponding convergence to the continuum limit is shown in Figure~\ref{oneloopcontlim}.
The deviations from the continuum phase diagram at $m_0>0$ and small $R/a$ are labeled as the ``lattice pert.\ regime'' in Figure~\ref{phasediag2-b}. In this regime the theory behaves rather similarly to QCD(adj) for $N_f>1$ in the continuum with a confined phase that extends up to larger $m$ the smaller $R/a$ is. The appearance of this regime is in agreement with the result of our first numerical simulations \cite{Bergner:2014dua}. 

The coefficient $\VR$ in the perturbative effective Polyakov loop potential of SYM provides some further indications for the difference between lattice and continuum phases: the numerical solution shows that it monotonically decreases with $R/a$. In the continuum $\VR=0$ at the perturbative level and due to the non-perturbative contributions it becomes an increasing function of $R$. There is hence a competition between non-perturbative effects and perturbative lattice artefacts: the transition to the ``lattice pert.\ regime'' is indicated by a turnover from the increasing continuum to the decreasing lattice perturbative behaviour, a minimum of $\VR$ and a maximum of the susceptibility. The investigation of the continuity in SYM is limited to a certain range of $R/a$ before this turnover, where the lattice artefacts are under control. This is the limit of large $R/a$ while $R$ is kept small. Even outside this regime, the qualitative properties are not completely different. They can be compared to $N_f>1$ QCD(adj) in the continuum and the qualitative features should be similar.

The perturbative continuum regime is reached considering the limits $\beta\rightarrow\infty$ and $R/a \rightarrow \infty$, the continuum limit for an asymptotically free theory regularized on the lattice. In this limit there is a simple perturbative relation between the renormalized mass $m$ and the bare mass in lattice units $m_0$. A regime where the perturbative effective potential provides a reasonable approximation is also approached in the small $R/a$ limit at fixed lattice spacing. However, this is not strictly perturbative, in particular a non-perturbative renormalization of the mass is required. The mass parameter $m_0$ in the perturbative formulas is replaced by $m/a$ fixed by some renormalization condition, in our case a fixed value of the adjoint pion mass in the large $R$ limit. The fixed scale approach that we apply in our current study allows to keep the line of constant physics, i.~e.\ constant renormalized parameters, unambiguously, compared to our previous studies in \cite{Bergner:2014dua}.

\subsection{Explanation of the discrepancy of lattice and continuum}
\label{explanation}
The phase diagram of lattice SYM with Wilson fermions and Wilson parameter set to $r=1$  exhibits a different phase structure than in the continuum limit.   In particular, for sufficiently small-bare mass, the lattice  theory is always confined and for a large mass, it is confining 
at large $R$, deconfined in an intermediate regime, and confined again at small-$R$, as shown in  Figure~\ref{phasediag2-b}.  
Note that the lattice regularization neither respects supersymmetry, nor chiral symmetry. In fact, if it were to respect one, it would respect the other as the only relevant chiral symmetry breaking operator $m \bar \lambda \lambda$ is also the only supersymmetry breaking  relevant operator for this system.  
 
 Here, we would like explain this discrepancy in slightly more detail, and argue that both of these conclusions are actually correct, and that there is no contradiction between  the two results.  
 
 Recall that both  the continuum and lattice  study aims to understand phase structure of the theory  by using 
  \begin{align}
 \tilde Z(R, m) = \tr \;  [e^{-R\hat{H}(m)} (-1)^F ] 
 \label{twisted}
 \end{align}
 In continuum, for $m=0$,   $\tilde Z(R, 0)$  is the supersymmetric Witten index and is independent of $R$. Hence, there is no phase transition on the $m=0$ axis for any value of $R$. In the small-$R$ regime,  
 there is no perturbatively induced potential for Wilson line. 
 There are  non-perturbative monopole-instantons with two fermion zero modes ${\cal M}_a = e^{-S_m}  e^{-  \vec \alpha_a \cdot (\vec \phi - i \vec \sigma) } (\vec \alpha_a  \lambda)^2$ where $\vec \phi $ is the holonomy field and   $\vec \sigma$ is the dual photon.   The crucial point is that despite the fact that monopole-instanton couple to holonomy field, it does not induce a boson potential as it has fermion zero modes. Rather, they do induce  a chiral condensate as well as a superpotential.
 Note that  $\prod_{a=1}^{N_c}  {\cal M}_a = I_{BPST}= e^{-S_I} (\tr \lambda \lambda)^{N_c}$, the usual 4d instanton. Therefore,   $ {\cal M}_a$ can be considered as fractional instanton. 
 The  second order effects in semi-classical expansion are of two types. Magnetic bions $[{\cal M}_a \overline {\cal M}_{a+1}]$ 
  and neutral bions $[{\cal M}_a \overline {\cal M}_{a}]$. In combination, these two type of bions induce mass gap for boson fluctuations (which are otherwise gapless to all orders in perturbation theory), and stability of center symmetry. 
 
The bosonic  potential for $\mathrm{SU}(2)$ SYM  is: 
 \begin{align} 
& V^{\rm pert.}  (\phi) = 0  \cr
&V^{\rm non.pert.}  (\phi) =  e^{- 2 S_m } \cosh \left( \frac{4\pi}{g^2} (2 \phi - \pi) \right)  \qquad S_m= {S_{\rm instanton} \over N_c } =  
\frac{8\pi^2}{g^2N}, 
 \end{align}
  which has a minimum at $\phi= {\pi \over 2}$.   Once a small mass for the fermion is added to the SYM action, 
  \eqref{twisted} is still a well-defined object, however  it is no longer  the Witten index as the supersymmetry is explicitly broken.  It  can be interpreted as a twisted/graded  partition  function in which boson and fermion states are added according to a 
  ${\mathbb Z}_2 =   (-1)^F $ grading.   
  
  Once a small-mass is turned on, there is a perturbatively induced GPY  potential, and monopole-instanton effects also induce a potential for the Wilson line holonomy as the soft mass lifts the fermion zero modes. Both of these effects are center-destabilizing and proportional to the soft-susy breaking parameter $m$, while the neutral bion effects are center-stabilizing. The competition between these three effects, neutral bions vs.\ perturbative and monopole-instanton  effects is responsible for the existence of the center and chiral phase transition. Also note that as $R$ gets smaller, the center breaking  effects certainly dominate, with no change for the center-symmetry to stabilize again. 
  
  On the other hand, what happens for lattice SYM is quite different. The phase diagram shown on  Figure~\ref{phasediag2-b}  is much like  the continuum analysis of the $1 < N_f < N_f^*$  theories. 
  The reason for this is following.  On the lattice, the dispersion relation of free bosons  and fermions are not equal.  The inverse propagator reads:
  \begin{align} 
{  \rm bosons:} & \qquad  \hat p^2 = \sum_{i=1}^4  4 \sin^2 \left(\frac{p_i }{ 2} \right)  \cr 
{ \rm Wilson \;  fermion:}  & \qquad  {\hat{\hat{p}}}^2  +  
\left( m_0 + \frac{r}{2}  \hat p^2  \right)^2 , \qquad    \hat{\hat{p}}^2 = \sum_{i=1}^4 \sin^2 p_i\, . 
  \end{align}
  If the Wilson term is set to $r=1$, indeed, there are no doublers but  rather interestingly, 
  a lattice Wilson fermion, in its effect to center-symmetry realization and to the potential \eqref{1loop}  is more powerful than a single continuum fermion! For this reason, the lattice theory with $N_f=1$ is behaving similar to $1 < N_f < N_f^*$  continuum theories. 
  
    In this sense, $N_f^{\rm  lattice}=1 $ is capable to undo the center-breaking effect of gauge fluctuations even at one-loop order.  This is the reason that center-symmetry is  stable at one-loop order in lattice perturbation theory in the $m_0 =0$ limit,
   \begin{align} 
& V^{\rm pert.}  (\phi) = N^2 \sum_{n=1}^{\infty} V_n |P_L^{(n)}|^2, \qquad  V_n >0.
 \end{align} 
 The lattice center-stabilizing effects given by Wilson fermions continue to hold true for sufficiently small bare mass $m_0$ of fermions.  Furthermore, for very large-$m_0$,  as one dials $R$, it is possible to show that 
$V_1$ is positive for $R < R_c$  and center symmetry is stable  and $V_1$ is negative for $R_c<R \lesssim  \Lambda$. This calculable phase transition correspond to the lower line in \ref{phasediag2-b}.     The upper line  is incalculable by weak coupling methods, but it is essentially the (inverse) deconfinement  temperature  of pure YM theory.  Note that fermions play 
a two fold role here. 
The interesting aspect is the { \it non-decoupling } of heavy fermions once the radius is taken small. In that regime, the fermion, despite being heavy, enters in the combination $m_0 R$ and sufficiently small $R$ makes the fermions behave as if they are light. When $R$ is sufficiently large,   the fermions essentially decouples from the dynamics and we have the transition of pure Yang-Mills theory. 

This observation can be phrased in a different way: at each $N_t=R/a$ there is a fixed value of $m_0=am$ for the transition. This means that as a function of $a$ in simulations at fixed $N_t$, the transition line follows a constant $m_c R_c$ like $N_f>1$ QCD(adj) in the continuum. This fixed $N_t$ approach has been applied in our previous simulations and the results show indeed a narrowing of the deconfined region towards smaller $a$, i.~e.\ larger $\beta$ \cite{Bergner:2014dua}. The continuum limit hence requires the large $N_t$ limit. At fixed $N_t=4$, lattice $N_f=1$ QCD(adj) is comparable to $N_f>4$ in the continuum, see Figure~\ref{fixednt}. Note that this simple relation holds only at the one loop level.

The fermion-boson mismatch of the dispersion relations is essentially  the reason why the lattice theory with $N_f^{\rm  lattice}=1 $ behaves differently from the continuum theory with $N_f=1$.  In the continuum limit, with the tuning towards supersymmetric point, the lower transition line in  Figure~\ref{phasediag2-b}  tends towards $R=0$ and one expects the continuum phase diagram shown in Figure~\ref{phasediag2-a}.  

In the next section, we present evidences from Monte-Carlo simulations giving rise to the phase diagram Figure~\ref{phasediag2-b}, 
providing a numerical proof of the idea of adiabatic continuity between the small and large radius for sufficiently 
light fermions. Furthermore, we will also provide evidence for the continuity of the phase transition lines between the weak coupling calculable regime and strong coupling incalculable regime.

\section{Numerical results for compactified adjoint QCD}
\label{sec:numres}
We have performed numerical simulations of SU(2) QCD(adj) with thermal and periodic boundary conditions for different fermion masses and couplings. Most of our simulation are for $N_f = 1$ but we have also considered $N_f = 2$ QCD(adj) in an exploratory study. The main focus are the Polyakov loop, its susceptibility, and the Binder cumulant, to investigate the theoretical conjectures about the deconfinement transition. 
In the fixed scale approach the compactification radius $R$ is proportional to the number of lattice sites in the temporal direction $N_t=R/a$. Since the lattice spacing is fixed, we can disentangle easily the renormalization effects and the tuning of the bare fermion mass from the change of the compactification radius. Based on our previous experience, a rather small lattice spacing is required since the lattice itself is a relevant source of supersymmetry breaking, even more relevant than the breaking induced by the gluino mass. For this reason we have compared different lattice actions.
\def\wid{.47}
\def\widd{.32}
\subsection{Confined and deconfined phases in SYM theory}
\begin{figure}
\begin{center}
\subfigure[\label{fig:pl16c-a}]{\includegraphics[width=\wid\textwidth]{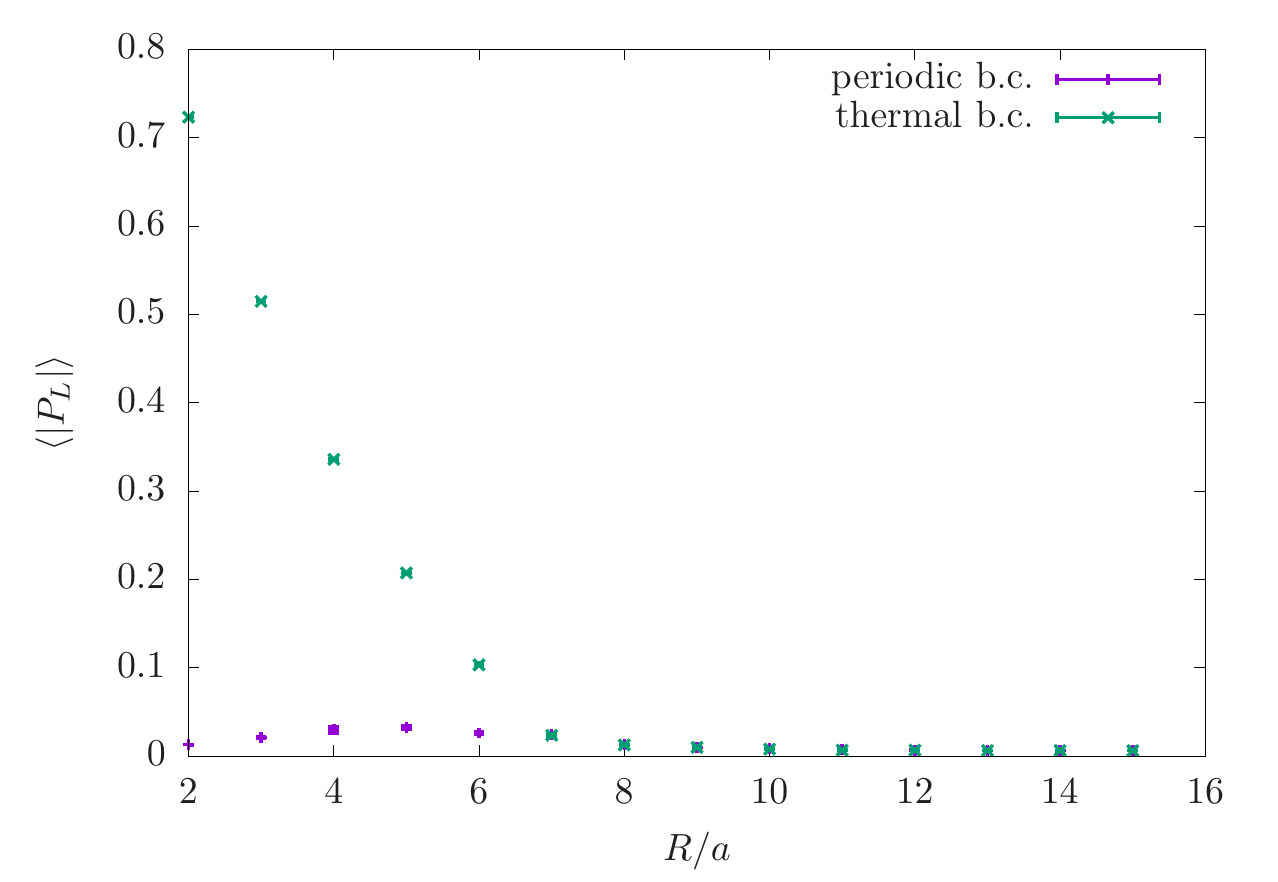}}
\subfigure[\label{fig:pl16c-b}]{\includegraphics[width=\wid\textwidth]{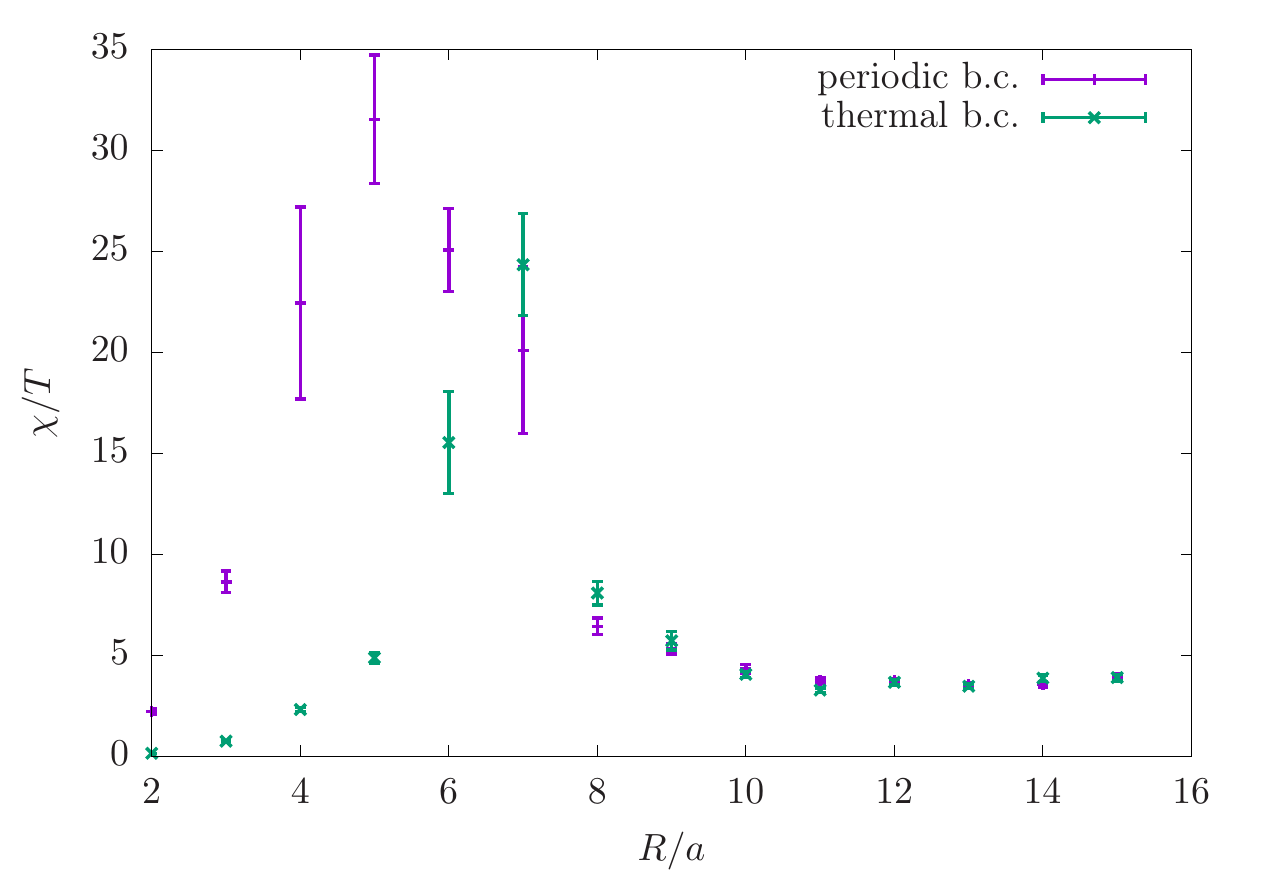}}
\subfigure[\label{fig:pl16c-c}]{\includegraphics[width=\wid\textwidth]{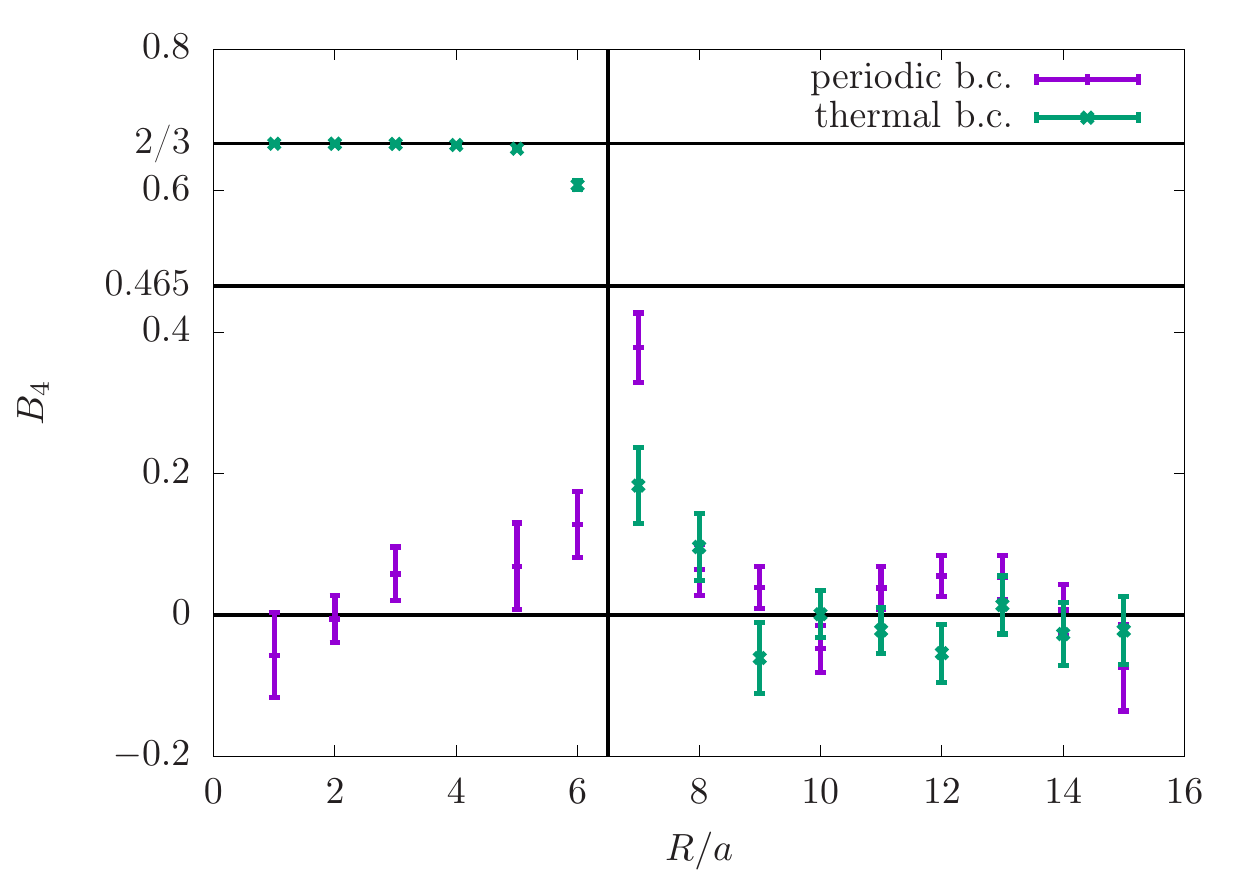}}
\end{center}
 \caption{
 The results for the Polyakov loop at $\beta=1.65$, $\kappa=0.1750$, $c_{sw}=1.0$ on a $16^3\times N_t$ lattice as a function of $N_t=R/a$. The plots show a comparison of periodic and thermal fermion boundary conditions:
 (a) the average modulus of the Polyakov loop $|P_L|$, (b) the susceptibility of $P_L$, and (c) the Binder cumulant. The  "pion mass" (gluino-ball)  at these parameters in lattice units is $am_\pi=0.64631(67)$.
 The vertical line in (c) indicates the point of the deconfinement transition.
 }\label{fig:pl16c}
\end{figure}
\begin{figure}
\begin{center}
\subfigure[\label{fig:plhist16c-a}]{\includegraphics[width=\widd\textwidth]{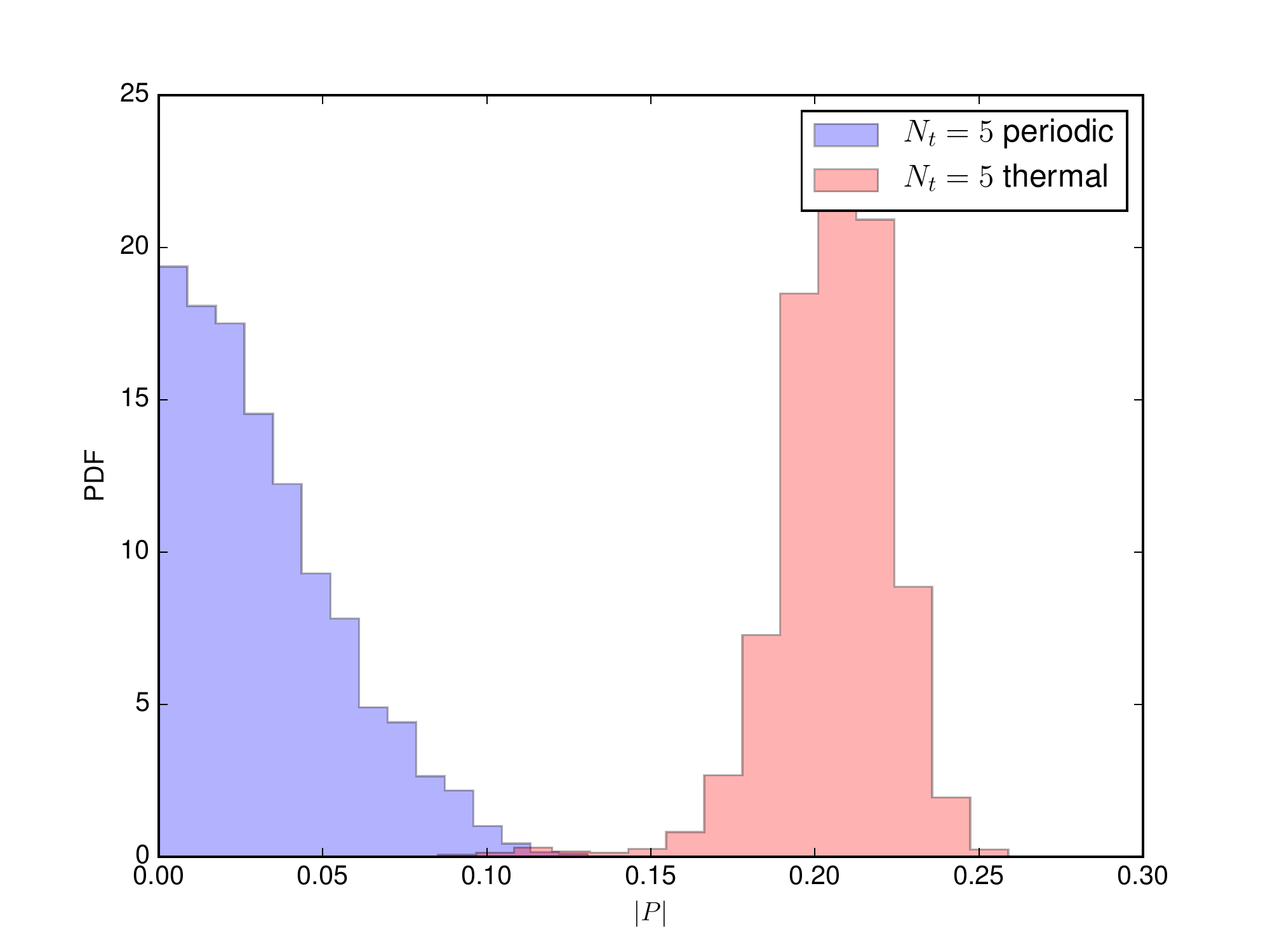}}
\subfigure[\label{fig:plhist16c-b}]{\includegraphics[width=\widd\textwidth]{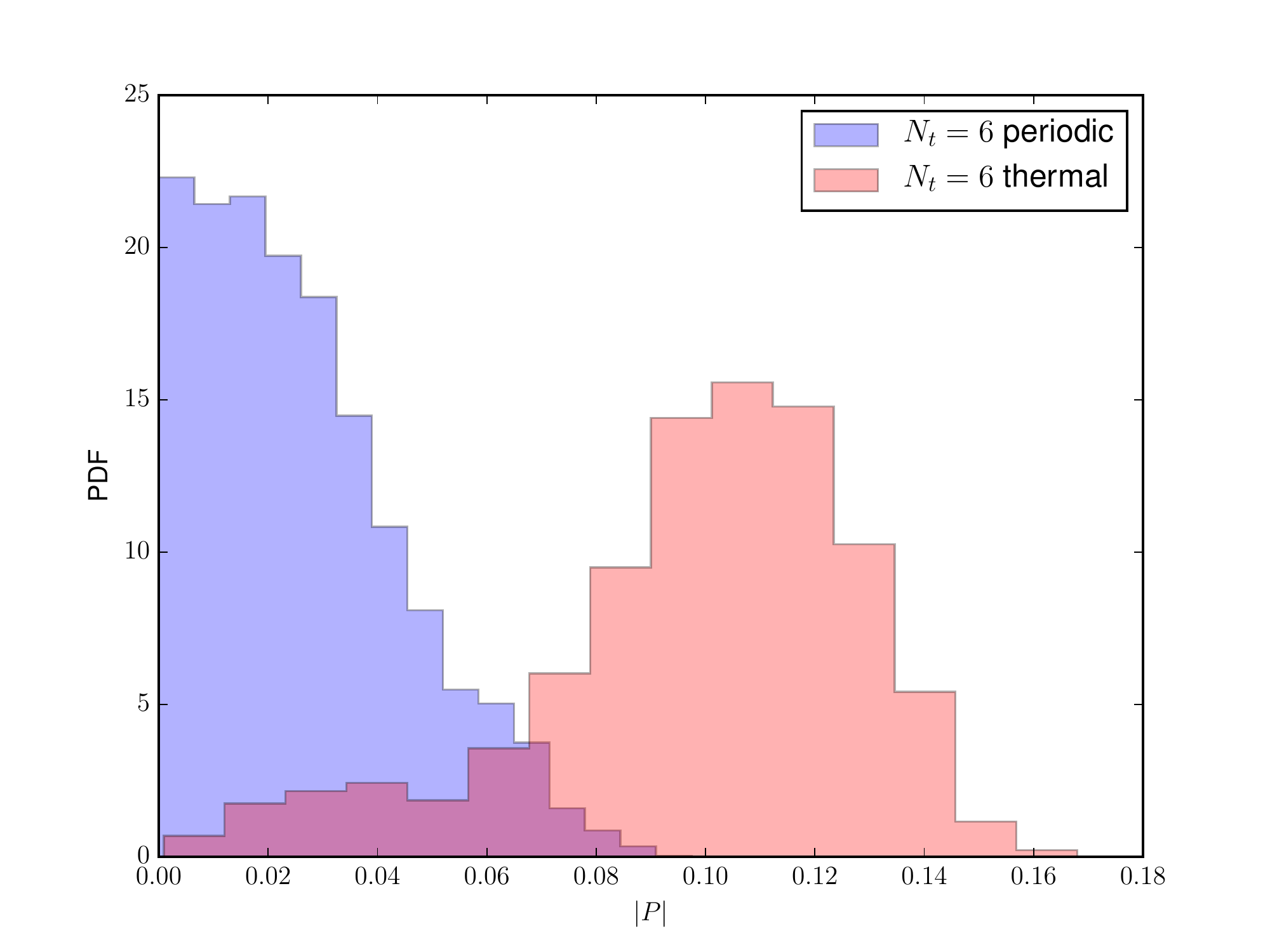}}
\subfigure[\label{fig:plhist16c-c}]{\includegraphics[width=\widd\textwidth]{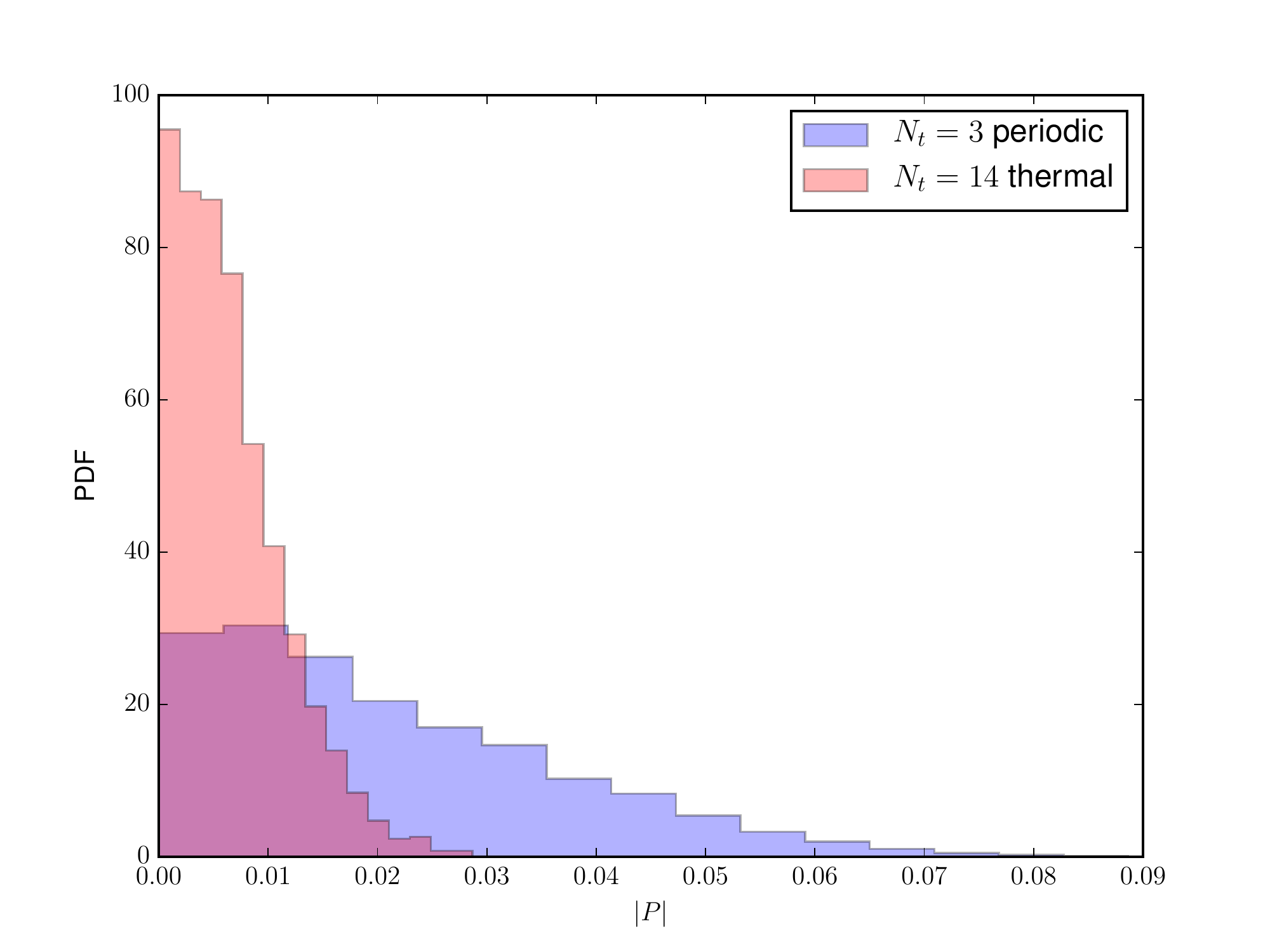}}
\end{center}
 \caption{The histograms for the Polyakov loop at selected $N_t=R/a$ for periodic and thermal fermion boundary conditions. The parameters
 are the same as in Figure~\ref{fig:pl16c}.
 }\label{fig:pl16cHist}
\end{figure}
\begin{figure}
\begin{center}
\subfigure[\label{fig:pl24c-a}]{\includegraphics[width=\wid\textwidth]{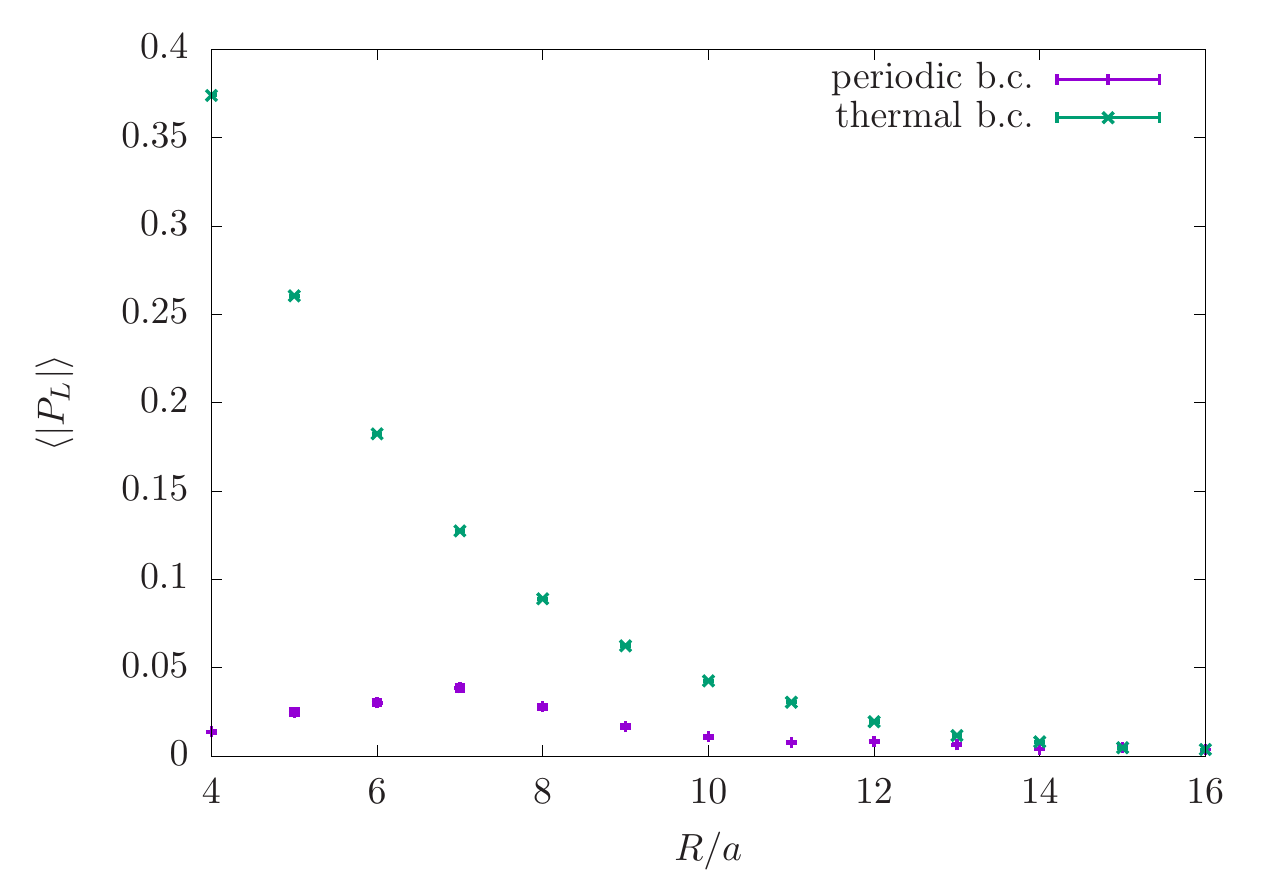}}
\subfigure[\label{fig:pl24c-b}]{\includegraphics[width=\wid\textwidth]{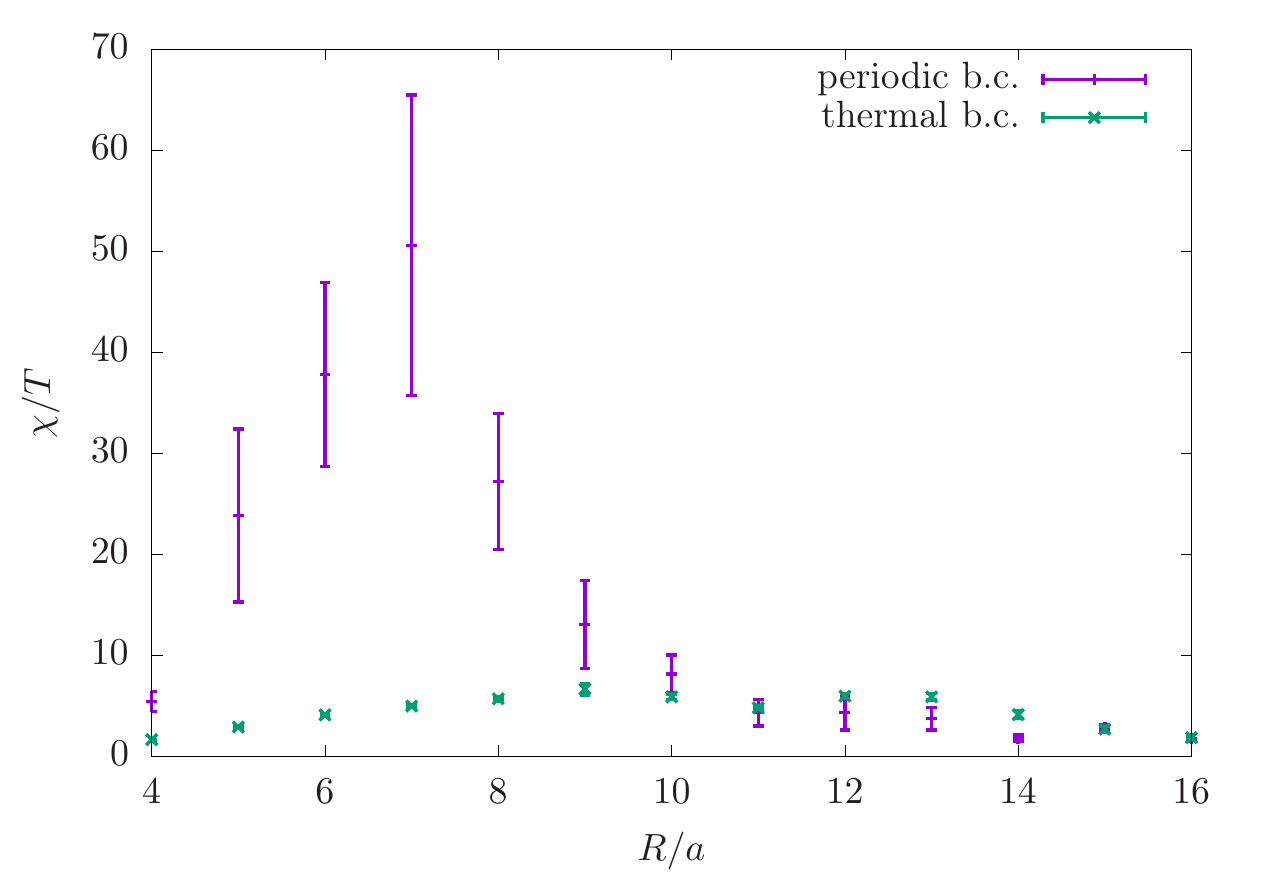}}
\subfigure[\label{fig:pl24c-c}]{\includegraphics[width=\wid\textwidth]{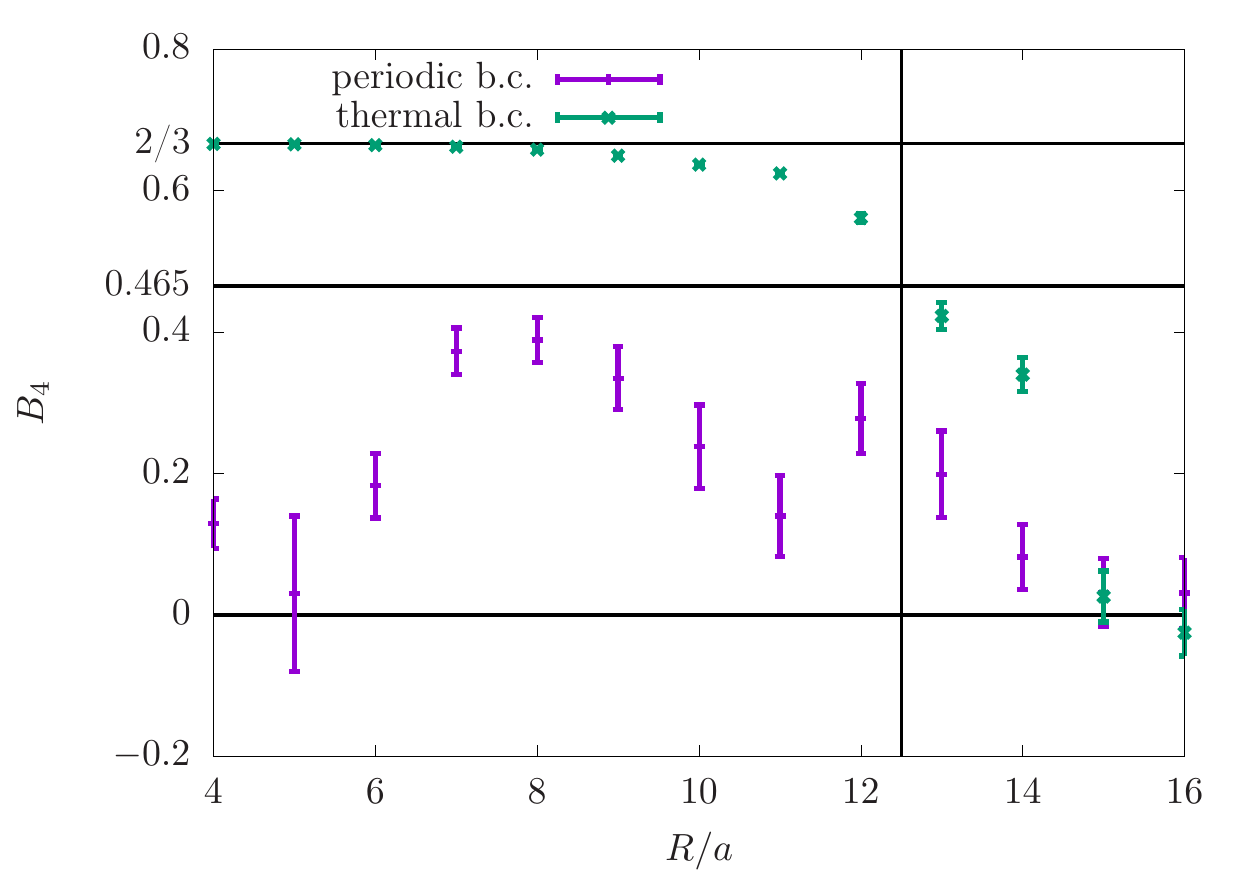}}
\end{center}
 \caption{
 The results for the Polyakov loop at $\beta=1.75$, $\kappa=0.14925$, $c_{sw}=0$ on a $24^3\times N_t$ lattice as a function of $N_t=R/a$. The plots show a comparison of periodic and thermal fermion boundary conditions in the same way as in Figure \ref{fig:pl16c}:
 (a) the average modulus of the Polyakov loop $|P_L|$, (b) the susceptibility of $P_L$, and (c) the Binder cumulant. In these simulations one level of stout smearing was applied to the gauge links in the Dirac operator.
 The physical observables at these parameters can be found in \cite{Bergner:2015adz}.
 }\label{fig:pl24c}
\end{figure}
The Polyakov loop and its susceptibility provide a signal for a possible deconfinement transition. Typically the transition can be identified by the peak of the susceptibility and by its divergence in the thermodynamic limit. Here a special care is required since the expected regime at small compactification radii is characterized by large fluctuations of the Polyakov loop, which are unrelated to the fluctuations between the confined and deconfined phases. The expectation value of $|P_L|$  and its susceptibility can become large without a deconfinement phase transition for periodic boundary conditions just because of the almost flat effective potential. A precise estimation of the phase of the theory can be obtained from the histogram of the Polyakov loop and from the Binder cumulant.

The Polyakov loop and its susceptibility are shown in Figure~\ref{fig:pl16c}.\footnote{\; \; In lattice simulations, it is a common practice to look at $\langle | P_L | \rangle$ instead of  $\langle  P_L  \rangle$. The reason for the absolute value is the fact  in finite volume and finite $N_c$,  there are no strict phase transitions. 
 The transitions between different  vacua which is  forbidden for  the theory on ${\mathbb R}^3 \times S^1$,  upon compactification down to  $T^3 \times S^1$,   is replaced by $e^{-{\rm Vol}(T^3)}$.   This indeed leads to spontaneous symmetry breaking as ${\rm Vol}(T^3) \rightarrow \infty$.   
If the simulation time is much  larger than the inverse tunneling probability, the Polyakov loop expectation value will average out to zero even in the deconfined phase. 
This will not take place if one uses  $\langle | P_L | \rangle$, giving a cleaner diagnostic of  confinement-deconfinement transition in finite volume. }
As expected, thermal boundary conditions lead to a deconfinement transition at $R/a\sim 7$ indicated by a steep rise of the Polyakov loop expectation value and by a peak of the susceptibility. The Polyakov loop expectation value for periodic boundary conditions remains rather small, but a large value of the susceptibility is observed at $R/a\sim 5$. 
However, even close to the maximum, the histogram of $P_L$ is still peaked at zero as in the confined phase. The large values of the susceptibility originate from a broadening of the Polyakov loop distribution but not from a phase transition, see Figure~\ref{fig:pl16cHist}.
 The constraint effective potential becomes nearly flat, but its minimum remains at zero.
At the smallest $R/a$ the susceptibility decreases again, a behaviour compatible with the predictions of lattice perturbation theory. 
The Binder cumulant shows further evidences for such a scenario. If thermal boundary conditions are chosen for the gluino, $B_4$ crosses the critical value of the $Z_2$ Ising universality class. Instead, the Binder cumulant never reaches this value in the periodic case, see Figure~\ref{fig:pl16c-c}, and it decreases toward zero at the smallest $R/a$, in agreement with the perturbative predictions.

We have simulated the theory also with a different lattice action and a different lattice spacing and obtained compatible results, Figure~\ref{fig:pl24c}. In this case a larger volume would be required for a clearer identification of the thermal deconfinement transition from the peak of the susceptibility. 
The transition at $T_c$  corresponding to $R/a\sim 12$ is more clearly identified by the Binder cumulant \ref{fig:pl24c-c}. The phase diagram is in good agreement with Figure~\ref{fig:pl16c}: for periodic boundary conditions very large peak value of the susceptibility is obtained at rather small compactification radius $R/a=R_p/a\sim 6$ and the susceptibility decreases at smaller $R$. The separation of the peak of the susceptibility for periodic boundary conditions from the one of the thermal transition becomes significantly larger at the smaller lattice spacing. This indicates that $R_pT_c$ becomes smaller and might even, as expected for continuity, vanish in the continuum limit. The absence of a confined phase in periodic compactified SYM is quite stable with respect to a change of the gluino mass.

\subsection{$N_f=2$ QCD(adj)}
\begin{figure}
 \centering
  \subfigure[]{\includegraphics[width=\wid\textwidth]{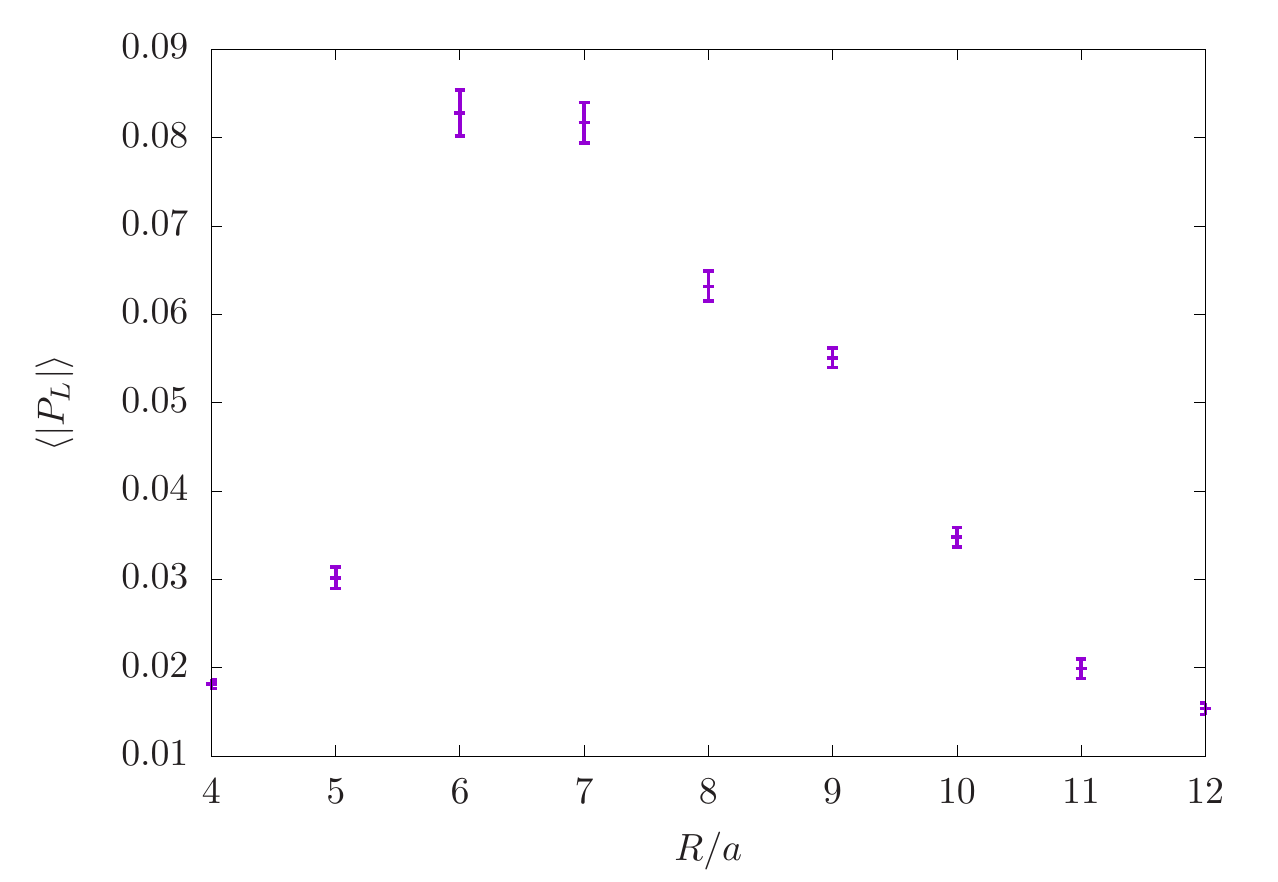}\label{fig:plnf2}}
 \subfigure[]{\includegraphics[width=\wid\textwidth]{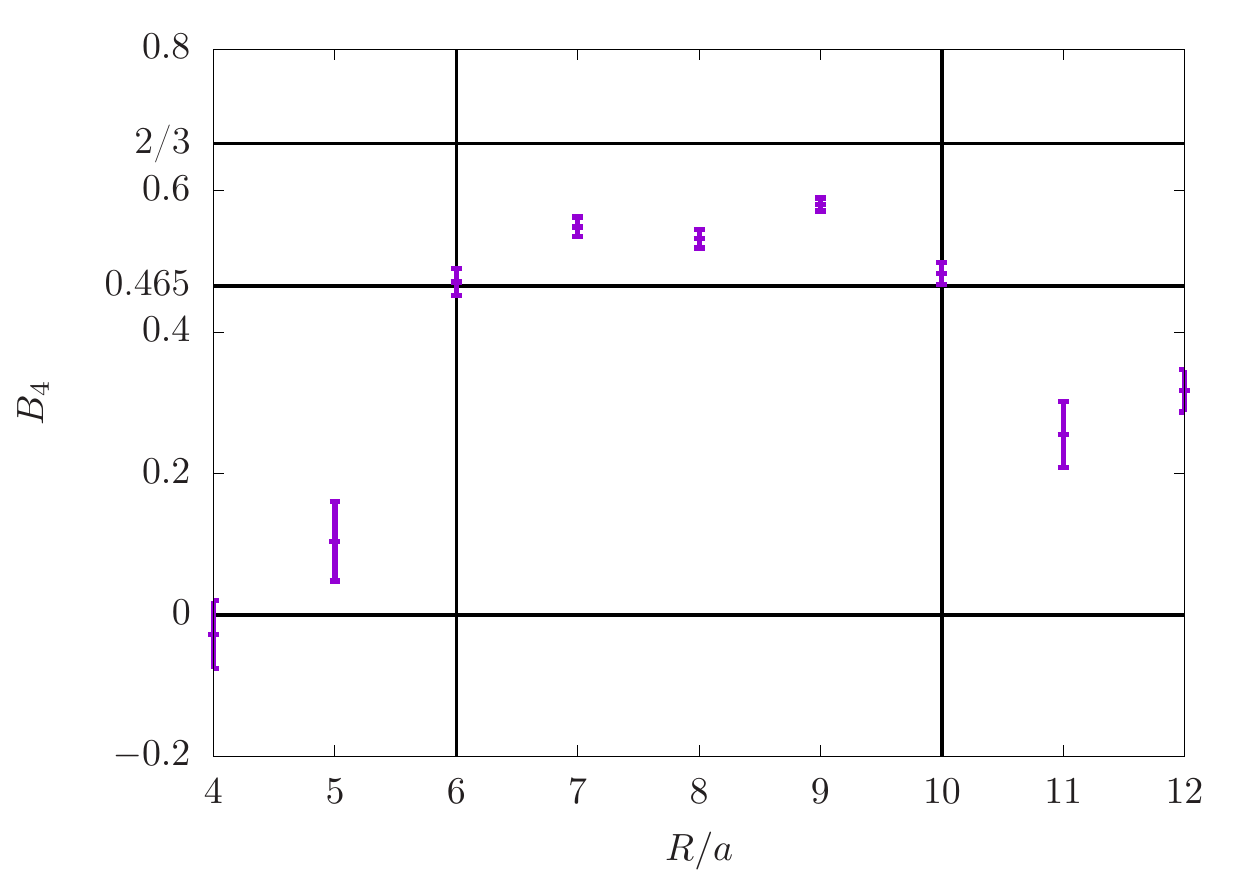}\label{fig:b4nf2}}
 \subfigure[]{\includegraphics[width=\wid\textwidth]{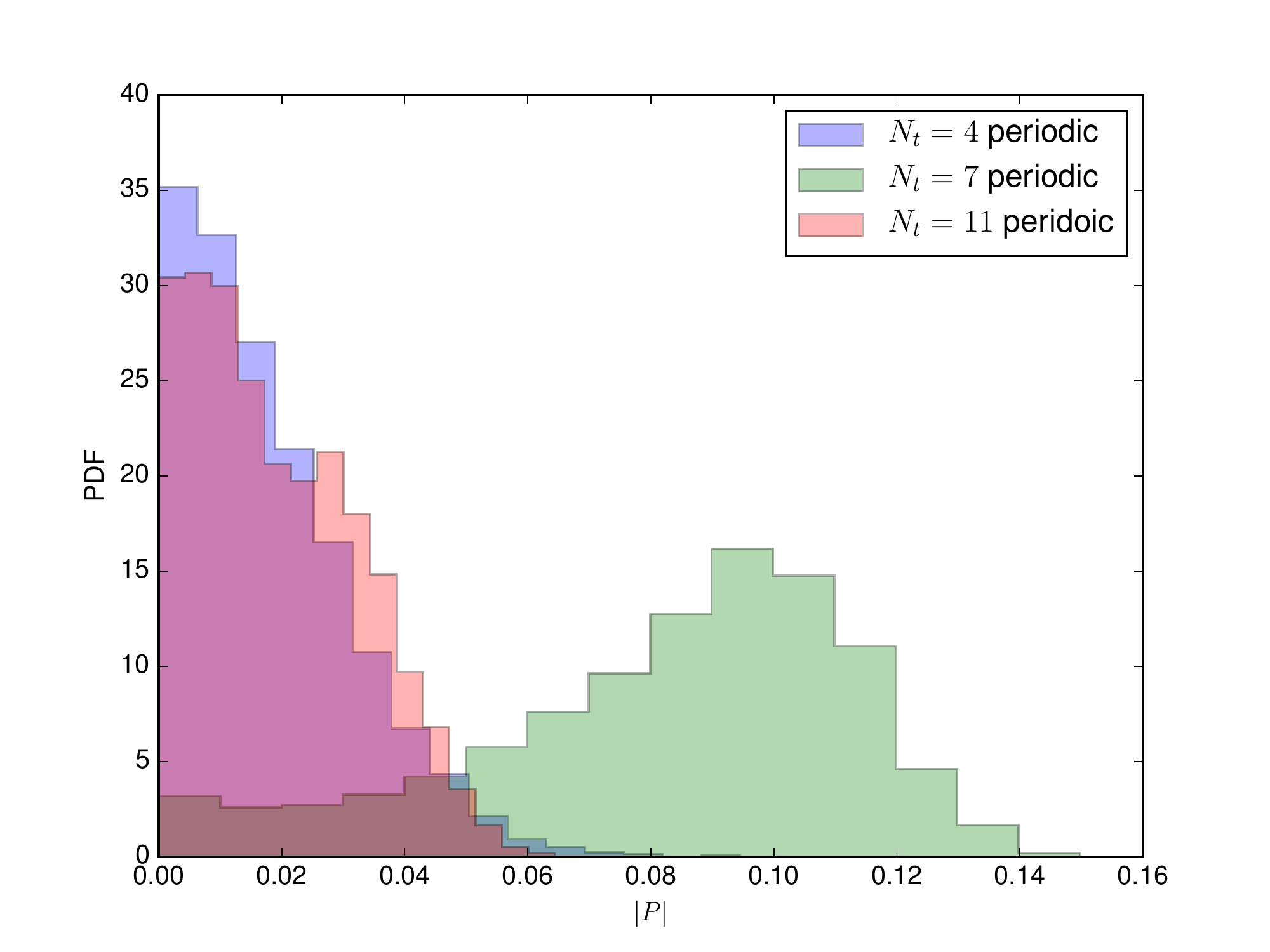}\label{fig:nf2hist}}
 \caption{Polyakov loop data from simulations of $N_f=2$ QCD(adj) at $\beta=1.75$, $\kappa=0.1620$ ($c_{sw}=1.0$) with periodic boundary conditions. (a) The average modulus of the Polyakov loop as a function of $R/a$, (b) the Binder cumulant, (c) the histogram of $|P_L|$. }\label{fig:nf2pldata}
\end{figure}
There is a limited knowledge about the properties of $N_f=2$ QCD(adj) at zero temperature. Recent numerical investigations have shown some evidence that the model could be already near the conformal window \cite{Athenodorou:2014eua}, see also \cite{Anber:2018iof} for a recent theoretical analysis. The running of the strong coupling has been found to be, however, quite different from zero \cite{Bergner:2017ytp}. The spectrum of the theory has a light $0^{++}$ glueball and an heavier pion(gluino-ball), while the opposite occurs in the QCD bound spectrum. This theory is free from the Pfaffian sign problem, but a near conformal behavior requires rather large volumes to be safe from finite volume corrections.

In the numerical simulations of this theory we have found a clear signal of a deconfined intermediate phase and confined phases at large and small $R/a$, see Figure~\ref{fig:nf2pldata}. The deconfinement can be clearly identified by the histogram in Figure~\ref{fig:nf2hist}.  The different theories can not be directly compared with each other, but in the simulations we have considered so far we never observed the disappearance of the intermediate phase. The intermediate phase is expected in general for heavy fermion masses, but for a conformal theory it should persist even for arbitrary low fermion masses, compare Figure~\ref{phasediag2-c}. It is hard to judge about the conformal behavior based on our present simulations, but at least they are not contradicting it.

The extension of the intermediate deconfined phase seems to be related to different competing effects. The one-loop calculation for Polyakov loop  suggests that the confined phase at small $R$ for a given $m$ becomes larger the larger $N_f$ is, which implies a shrinking deconfined phase with $N_f$. Fermions enforce center-stability and confinement in the perturbative small $R$ regime. On the other hand, fermions induce screening effects in the $\beta$ function. In the chiral limit, 
the leading order coefficient of the beta function is $\beta_0= \left(\frac{11}{3} - \frac{2}{3}{N_f} \right)N$. Therefore, fermion fields 
also  counteract confinement at large $R$ due to screening effects.  Our results tend to indicate that the sum of these effects prefers the extension of the deconfined phase, but further analysis and a more detailed consideration of finite volume effects are required in order to make a clearer statement about the continuity at $N_f=2$. Further studies of the two-flavor theory might be interesting in particular regarding the conjectures of \cite{Anber:2018iof} about the chiral transition. 

\subsection{Eigenvalue distributions and Abelian vs. non-abelian confinement}
\label{sec:effpot}
\begin{figure}
 \centering
 \includegraphics[width=.8\textwidth]{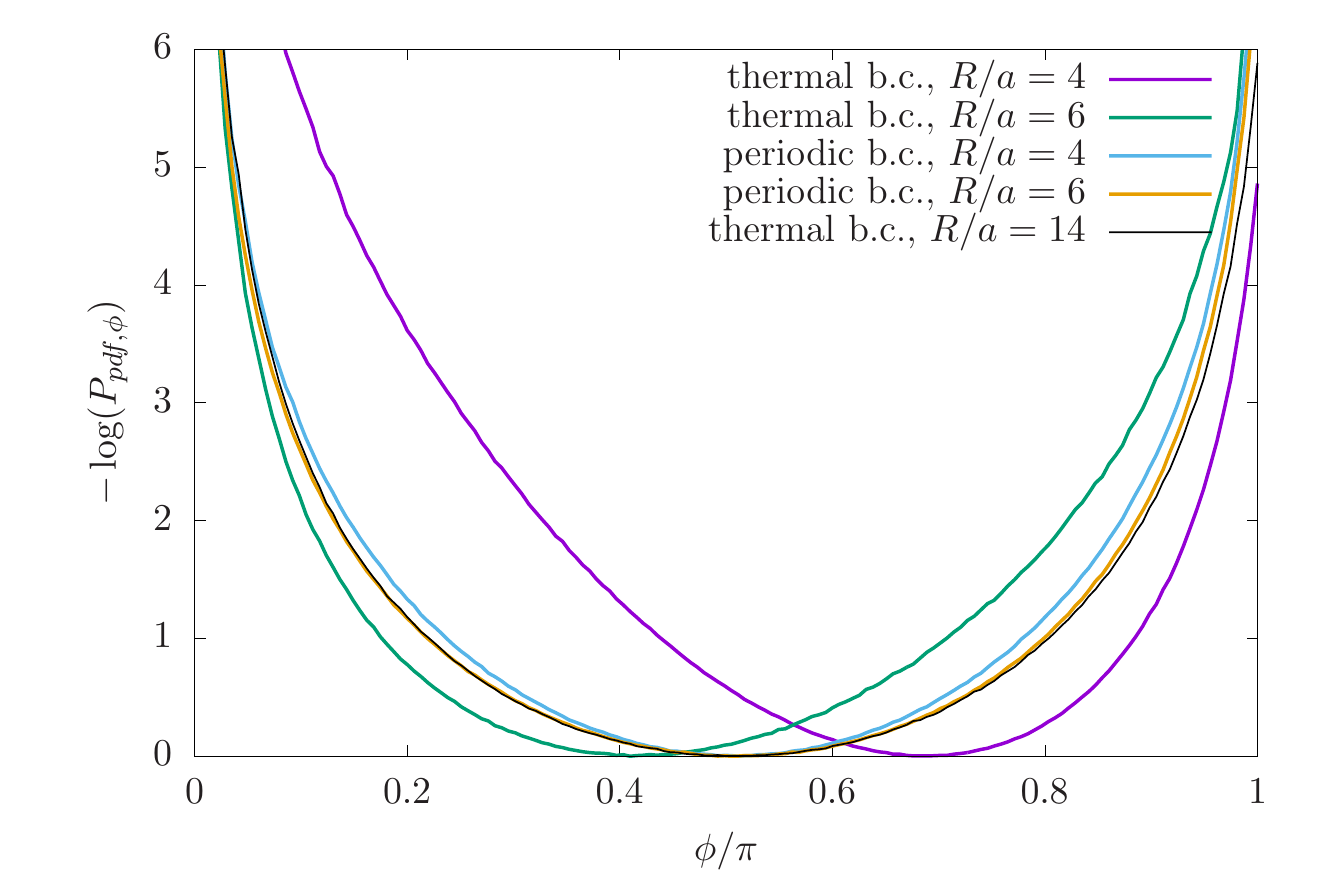}
 \caption{Constraint effective potential from the per-site distribution of the Polyakov phase $\phi$ comparing periodic and thermal fermion boundary conditions. The minimum at $\phi>\pi/2$ and $\phi<\pi/2$ for $N_t=4$ and $N_t=6$ in the thermal case indicates the negative and positive average value of the Polyakov line in the center broken phase. The histogram has been determined the same runs as in Figure~\ref{fig:pl24c}. The plot in this representation can be directly compared to the pure gauge case \cite{Smith:2013msa}.\label{fig:logpersitedistribution}}
\end{figure}
\begin{figure}
 \centering
 \includegraphics[width=.8\textwidth]{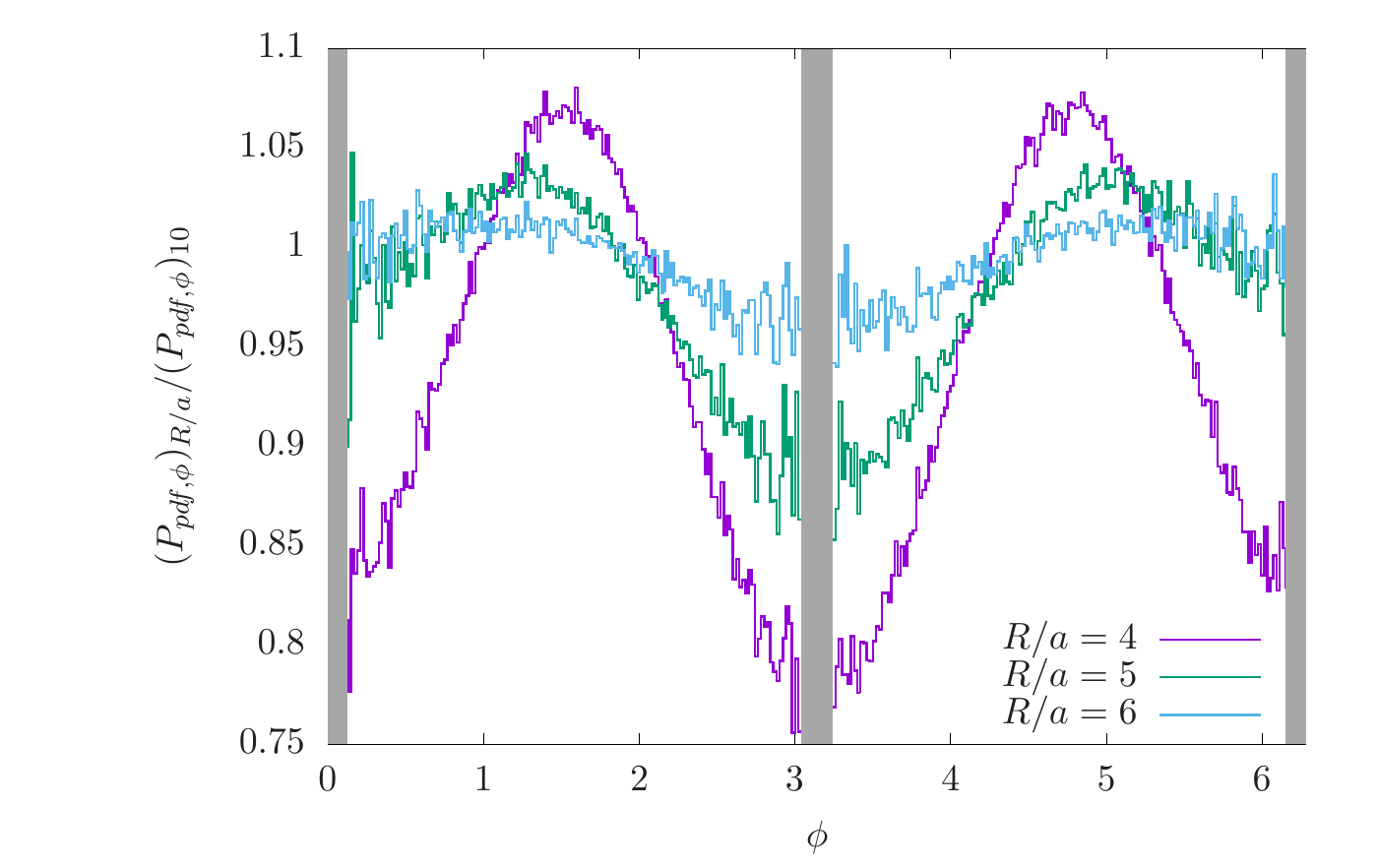}
 \caption{Per-site distribution of the Polyakov phase $\phi$ for different $R/a$ normalized with respect to the distribution at $R/a=10$, which is close to the Haar measure. 
 The histogram has been determined the same runs as in Figure~\ref{fig:pl24c}.
 The normalization of the histogram generates large errors in the regions close to $0 \mod \pi$, corresponding to the gray bands in the plot. These regions have been excluded.\label{site_eigenvalue_distribution_norm}}
\end{figure}
\begin{figure}
 \centering
 \subfigure[]{\includegraphics[width=.47\textwidth]{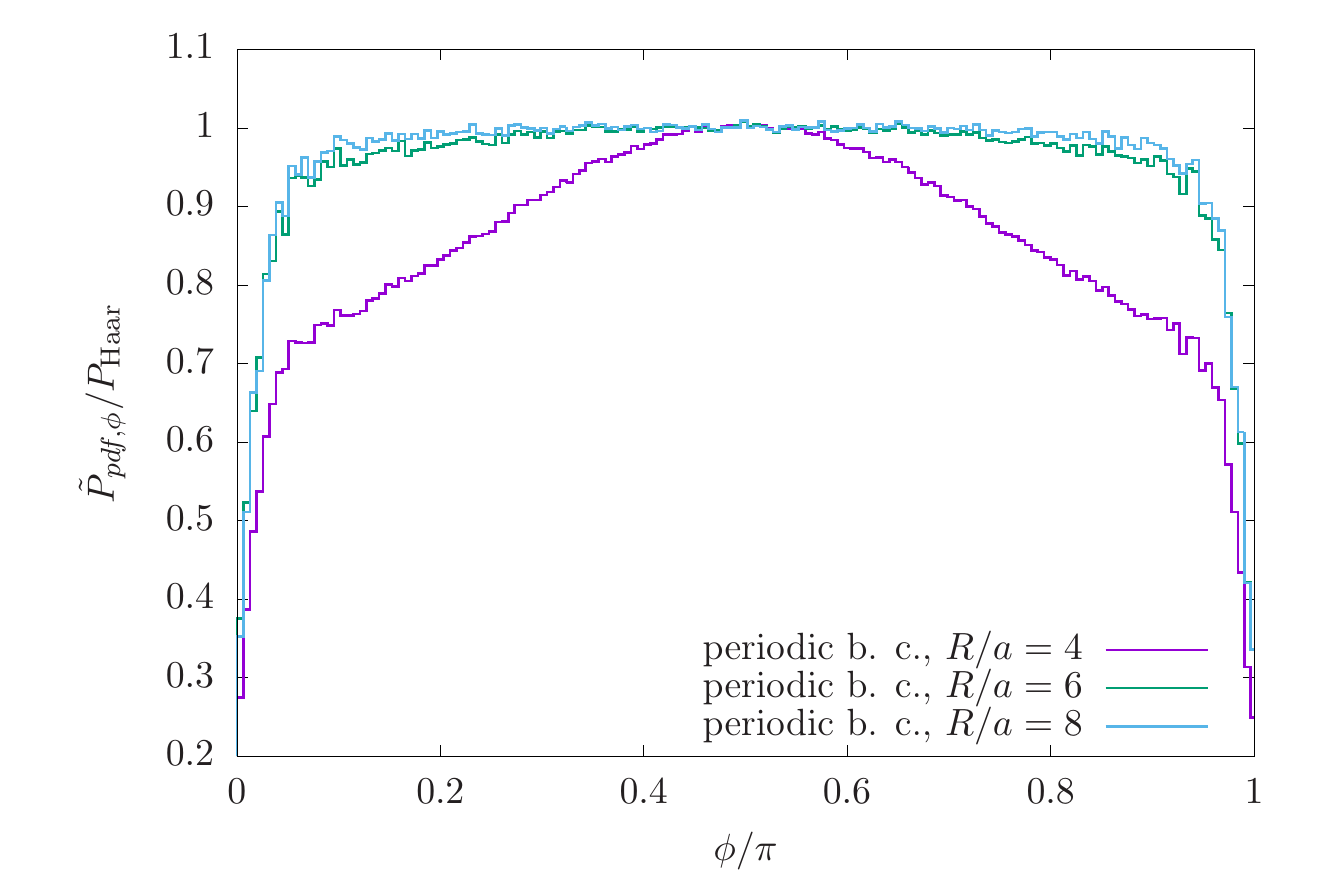}}
 \subfigure[]{\includegraphics[width=.47\textwidth]{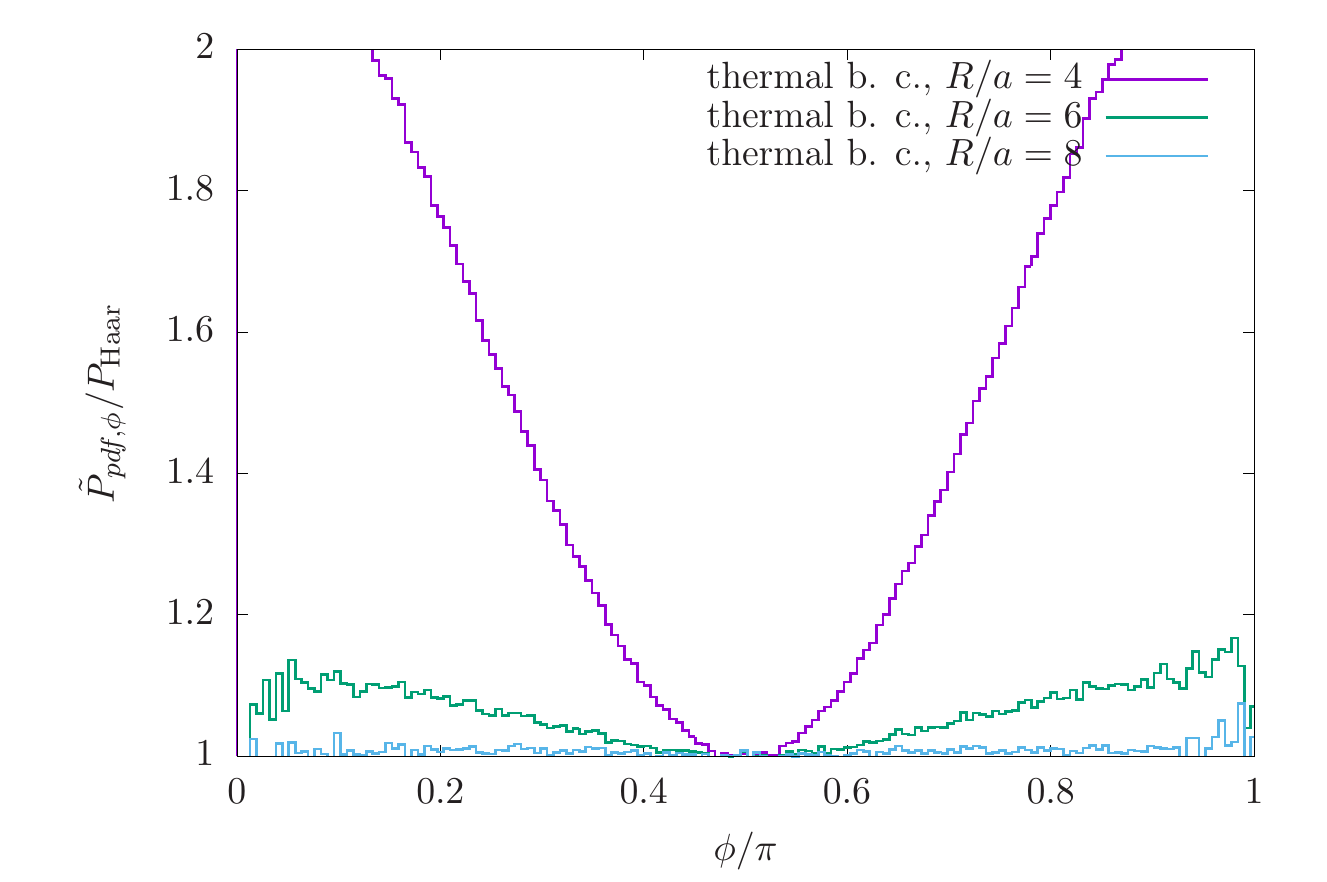}}
 \caption{Symmetrized per-site distribution of the Polyakov phase $\phi$ normalized with respect to the Haar measure for different $R/a$. 
 The histogram has been determined the same runs as in Figure~\ref{fig:pl24c} for periodic (Figure (a)) and thermal (Figure (b)) fermion boundary conditions.
 \label{fig:symeigdistr}}
\end{figure}
In this section we consider the distribution of the Polyakov line eigenvalues in SYM with the same parameters as in Figure~\ref{fig:pl24c}. As discussed in Section~\ref{sec:plactions}, the per-site distribution corresponds to the per-site constraint effective potential. It is different from the effective action of the volume averaged Polyakov eigenvalue phase and in the strong coupling limit it becomes equivalent to the Haar measure. We consider the probability function $P_{pdf,\phi}(\phi)$ of finding a Polyakov line at an arbitrary spacial point ${\bf x}$ whose eigenvalues are $\exp{(\pm i \phi({\bf x}))}$ in a random configuration of the Markov chain. 

\vspace{3mm}
 {\bf Abelian vs. non-Abelian confinement:}
The constraint effective potential obtained from the distributions is shown in Figure~\ref{fig:logpersitedistribution}. In case of thermal boundary conditions it is similar to the pure gauge case \cite{Smith:2013msa}. The asymmetric form above the deconfinement transition indicates a positive or negative expectation value of the Polyakov line.  
In the periodic case the plain probability function does not show a considerable dependence on the compactification radius $R/a$.  The deviations become much clearer if the per-site distribution is normalized by the Haar measure, see Figure~\ref{site_eigenvalue_distribution_norm}.  Due to numerical instabilities, the division by the Haar measure produces some artificial distortion in the regions where it is small.  There are two possible mechanism which may render $\langle P_L\rangle=0 $  in SYM compactified on a circle and both of these are taking place in the theory, albeit at different regimes, that we refer to as non-Abelian and Abelian confinement. 
\begin{itemize}
{\item  {\bf Random distribution:}  The flat distribution observed at large compactification radii is compatible with the non-Abelian confinement picture: random fluctuations of the phase $\phi$ in a nearly flat effective potential are the origin of the a vanishing expectation value for the Polyakov loop. } 
\item  {\bf Abelianizing distribution:}
At small $R/a$ the behaviour might be identified with the picture of Abelian confinement, where confinement is produced by a distribution of $\phi$ centered around the zeros of $\cos(\phi)$.  
\end{itemize}
It is remarkable that the theory up to a very small radius $R\sim 2/T_c$ shows a distribution compatible with the strong coupling limit.\footnote{We thank Misha Shifman for discussions on random vs. abelianizing distributions and on the connections of non-Abelian vs. Abelian confinement. }
SYM, and more general QCD(adj), as well as center-stabilized Yang-Mills and QCD theories  are expected to possess both regimes, an Abelian confinement regime where adjoint Higgsing is operative, and a strong coupling regime where random fluctuation of the phase of the Polyakov loop take over. Our numerical results provide evidence for the continuity between these two regimes.

Any constraint effective potential obtained from the distribution of the order parameter is symmetric under center symmetry on a finite volume in the limit of infinite statistic. The spontaneous breaking of center symmetry appears in the constraint effective potential of the volume averaged order parameter in terms of two minima related by the broken symmetry. The barrier between these minima diverges in the infinite volume limit and this signal is stable under a symmetrization of the distribution. Such an effect does not appear in the per-site distributions. With a finite statistics and spontaneously broken center symmetry the non-zero average Polyakov loop will lead to asymmetric distributions as in  Figure~\ref{fig:logpersitedistribution}, but a symmetrized distribution will in general not show a double peak. Nevertheless it is instructive to investigate the symmetrized per-site distributions in order to qualify the difference to the Haar measure. Symmetrization of the distributions with respect to center symmetry means $\tilde{P}_{pdf,\phi}(x)=\frac{1}{2}(P_{pdf,\phi}(x+\pi)+P_{pdf,\phi}(x))$ for SU(2). The symmetrized distributions normalized to the Haar measure are shown in Figure~\ref{fig:symeigdistr}. The difference between the preference of the confined or deconfined minima compared to the Haar measure can be clearly observed.

Information about the distribution of the volume averaged $\phi$, which is more directly related to the effective potential, can be obtained from the susceptibility of this observable. Like for the case of the traced Polyakov line in Figure~\ref{fig:pl24c}, there is a peak at a certain $R/a$ that indicates the flattening of the effective potential, but not a phase transition. It is expected that the maximum of the susceptibility moves to smaller and smaller $R/a$ in the continuum limit and the effective potential becomes flat (up to non-perturbative contributions) at small $m$ and $R$. 

Different from the expected continuum behaviour, we observe at very small $R/a$ at the same time decreasing susceptibilities of $P_L$ and $\phi$ together with a narrowing of the per-site distribution divided by the Haar measure. This is an indication of the ``lattice pert.\ phase'' in Figure~\ref{phasediag2-b} with suppressed fluctuation of $\phi$. It is a lattice artefact, but not a severe one since the theory becomes just similar to QCD(adj) at larger $N_f$. The region where regarding the vanishing perturbative contributions compactified lattice SYM is most similar to continuum SYM at small $R$ is close to the maximum of the susceptibility.

\subsection{The Polyakov loop in the adjoint representation}
\begin{figure}
 \centering
 \includegraphics[width=.57\textwidth]{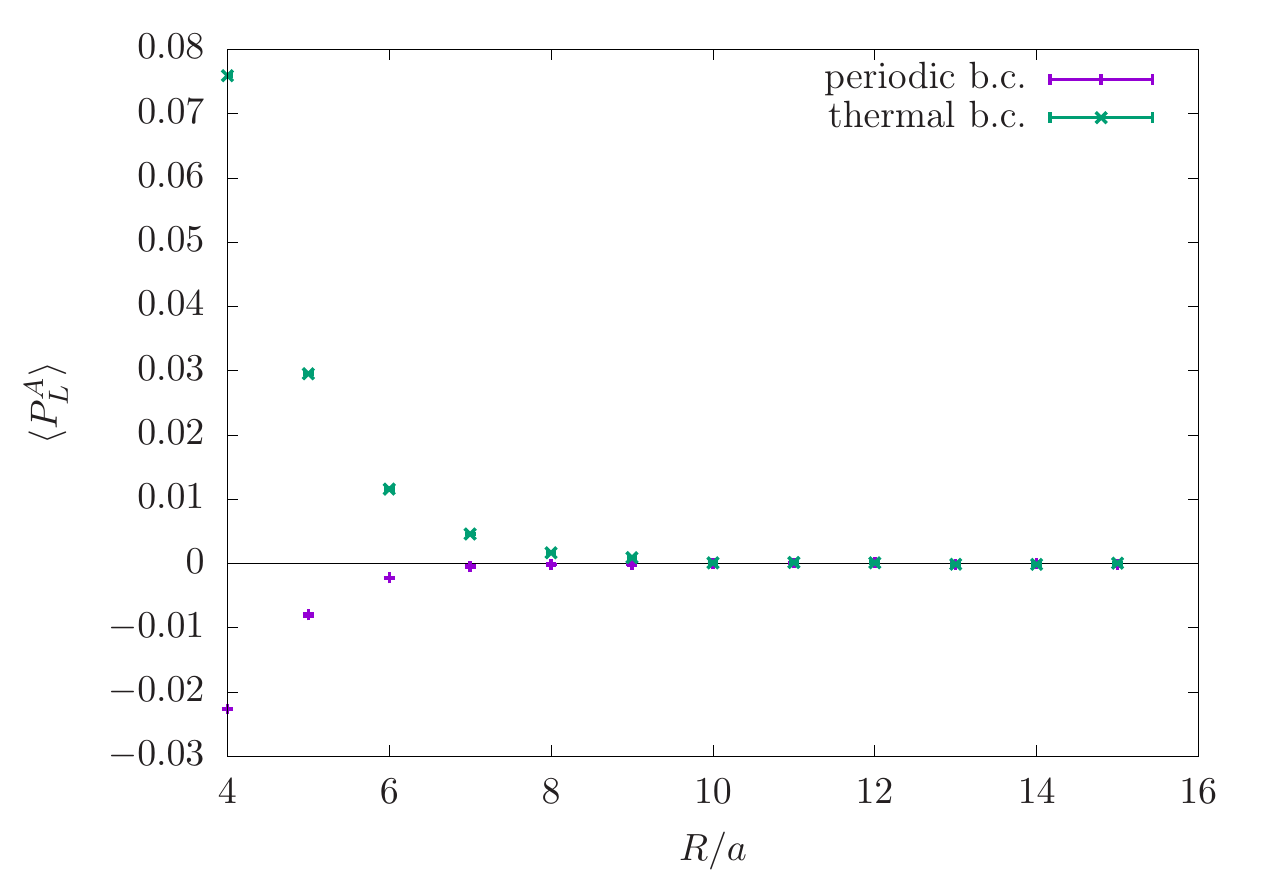}
 \caption{Polyakov loop in the adjoint representation as a function of the compactification radius for periodic boundary conditions with the same parameters as in Figure~\ref{fig:pl24c}.}\label{polyakov_loop_adjoint}
\end{figure}
In addition to $P_L$, we have also measured the expectation value of the Polyakov loop in the adjoint representation. Even at zero temperature, it is expected to have a non-vanishing expectation value since it is not fixed by center symmetry. 
It is interesting to note that for periodic boundary conditions the adjoint Polyakov loop develops a negative expectation value at small $R/a$, while the Polyakov loop in the fundamental representation is always close to zero, see Figure~\ref{polyakov_loop_adjoint}.
The adjoint Polyakov loop for thermal boundary condition develops, on the other had a positive expectation value.

The non-vanishing expectation value of the adjoint Polyakov loop can be approximately related to perturbative effective potential and the behavior of the per-site distribution of $\phi$. On each site the phase $\phi({\bf x})$ determines the trace of the fundamental Polyakov line as $\cos(\phi)$ and the adjoint one as $1 + 2 \cos(2 \phi)$. One can assume that the expectation values are basically determined by the per-site distribution of $\phi$:
\begin{eqnarray}
 \langle P_L \rangle & \approx & \frac{1}{2}\int_0^{\pi} 2 \cos(\phi) P_{pdf,\phi}(\phi) d \phi \,,\label{pfund}\\
 \langle P^A_L \rangle & \approx & \int_0^{\pi} (1 + 2 \cos(2 \phi)) P_{pdf,\phi}(\phi) d\phi\label{padj}\,.
\end{eqnarray}
The probability distribution function $P_{pdf,\phi}(\phi)$ at strong coupling (larger $R/a$) is proportional to the Haar measure. In this case both integrals are zero, meaning a vanishing expectation value of both in the adjoint and in the fundamental representation. 
At smaller $R/a$, the probability distribution found in Section~\ref{sec:effpot} can be well described by a functional of the form
\begin{equation}
 K -\cos(2\phi) + \alpha \cos(4\phi)\,,
\end{equation}
where $\alpha$ is a small positive constant, $0 \le \alpha < 0.5$. The term proportional to $\cos(4\phi)$ still vanishes in the integral \eqref{pfund}, but the integral \eqref{padj} is now equal to
\begin{equation}
 \langle P_A \rangle = \frac{\alpha}{\alpha-1} < 0\,.
\end{equation}
This implies again a negative value of the adjoint Polyakov loop. However, this is only a crude approximation that becomes exact only in the strong coupling limit.
The main effect is due to a localization (Abelianization) of the distributions: the random distribution of $\phi$ according to the Haar measure leads to $\langle P_A \rangle=0$, $\langle P_L \rangle=0$. 
In the completely localized limit, on the other hand, $\phi$ can be set to the perturbative minimum $\phi=\frac{\pi}{2}$ for periodic and  $\phi=0$ for thermal boundary conditions. 
In this saddle point approximation, the values are $\langle P_A \rangle\neq 0$, $\langle P_L \rangle= 0$ for the periodic and $\langle P_A \rangle\neq 0$, $\langle P_L \rangle\neq 0$ for the thermal case.
SYM in the continuum has only a non-perturbative localization of the distributions. Lattice artefacts provide localization effects even at the perturbative level, but still the numerical results show only a 
small non-zero value of $\langle P_A \rangle$ at the smallest $R/a$.

 \subsection{The chiral condensate}
 \begin{figure}
 \centering
 \includegraphics[width=.57\textwidth]{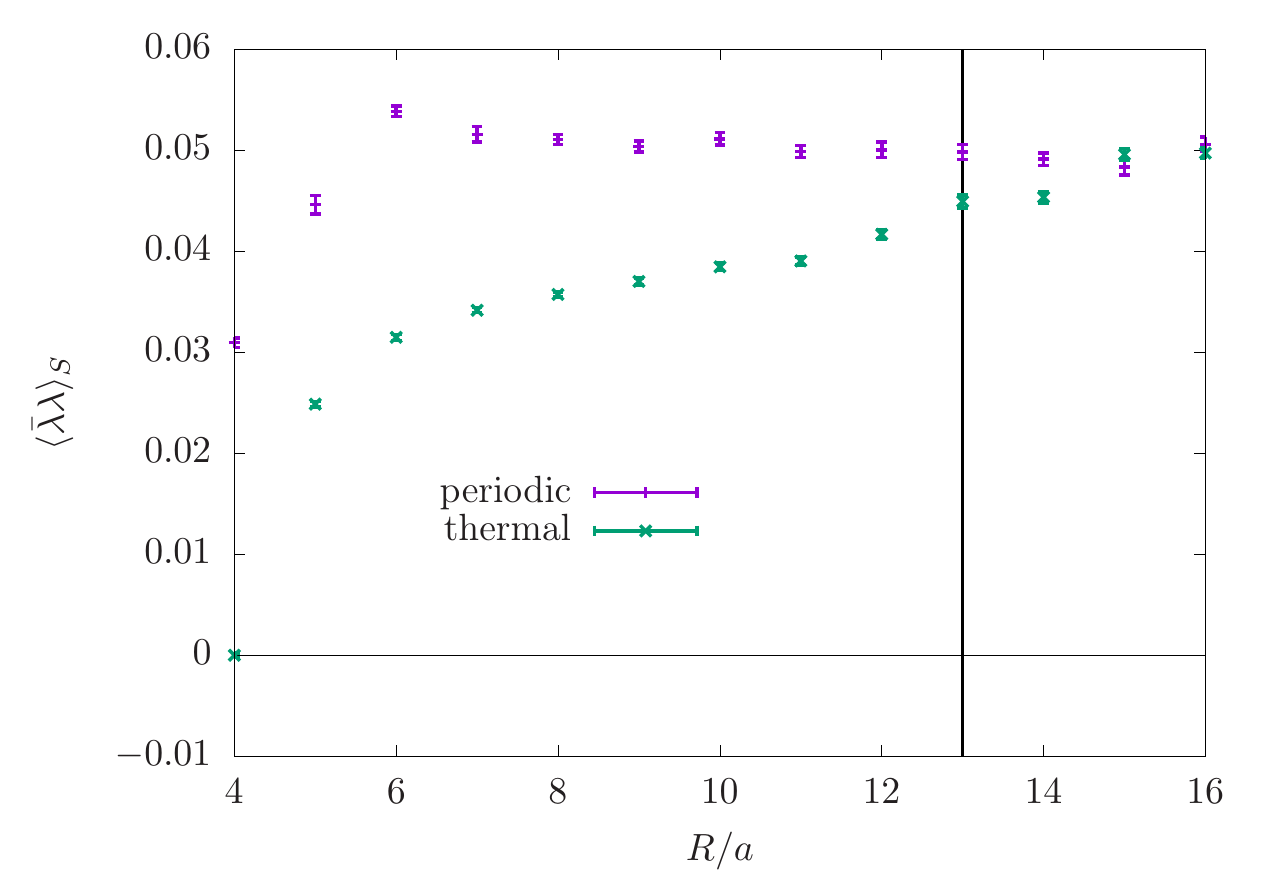}
 \caption{Subtracted chiral condensate as a function of the compactification radius for periodic and anti-periodic boundary conditions with the same parameters as in Figure~\ref{fig:pl24c}. The black vertical line indicates the maximum of the chiral susceptibility for anti-periodic boundary conditions. Note the persistence of the chiral condensate when fermions are endowed with periodic boundary condition up to much smaller circle size.}\label{chiral_condensate_periodic_comparison}
\end{figure}
The chiral condensate provides information about the spontaneous chiral symmetry breaking as a function of $R/a$. The bare condensate corresponds to the derivative of the partition function with respect to $m$,
\begin{equation}
\langle \bar{\lambda} \lambda \rangle_B 
\doteq -\frac{1}{RV_3}\frac{\partial}{\partial m} \log(Z(\beta,m)).
\end{equation}
A rough one-loop perturbative prediction can be computed from a derivative of the effective potential with respect to $m$ setting $\phi$ to the value of the minimum, corresponding basically to the trace of the fermion propagator in the given gauge-field background. In this perturbative limit one obtains rather small values that are decreasing with $R/a$ and proportional
to the mass term including the Wilson mass.

Recall that since Wilson fermion does not respect chiral symmetry, the condensate receives both additive and multiplicative renormalization. The subtracted condensate,
\begin{equation}
\langle \bar{\lambda} \lambda \rangle_S =   \langle \bar{\lambda} \lambda \rangle_B^{R} -  \langle \bar{\lambda} \lambda \rangle_B^{R=R_{r}} \;,
\end{equation}
 is considered to remove the additive renormalization.
The subtraction is performed with respect to a  fixed radius $R_r$, in the present case $R_r/a=4$ and thermal boundary conditions. The transition from the low temperature broken phase with a non-vanishing expectation value of the condensate to the unbroken phase is indicated by a jump of $\langle \bar{\lambda} \lambda \rangle_S$. The subtracted condensate as a function of $N_t=R/a$ is shown in Figure~\ref{chiral_condensate_periodic_comparison} for periodic and thermal boundary conditions. In the thermal case there is a signal for a chiral transition around the same point as the deconfinement transition. It can be identified by the peak of the chiral susceptibility. The considerable decrease  of the thermal $\langle \bar{\lambda} \lambda \rangle_S$ at very small $N_t$ is an indication of lattice artefacts introduced by the Wilson fermion discretization. The chiral susceptibility remains small in this region, which indicates the absence of a physical phase transition. 

The subtracted chiral condensate for periodic boundary conditions is close to  its large-$R$ value up to a small radius $R/a\sim 6$.  At an even smaller radius there is an decrease  of the subtracted chiral condensate in accordance with the perturbative prediction. This appears where also in the case of thermal boundary conditions, the lattice artefacts introduce a significant distortion. The decreasing values of the chiral condensate might also indicate a relation to the adjoint Polyakov loop such as found in  \cite{Gattringer:2006ci}. Our findings can also be compared to the investigation of the chiral order parameter with adjoint staggered fermions at larger $N_f$ and periodic boundary conditions in \cite{Cossu:2009sq}, where it was found that the chiral symmetry remains spontaneously broken to small $R$ with an indication for a chiral symmetry restoration at very small $R$.

The crucial point  is that unlike thermal compactification, the chiral condensate persists up to rather small circle size in the circle compactication.  In fact, as long as center-symmetry is intact, chiral symmetry breaking remains robust.  This provides further evidence for the idea of adiabatic continuity.

\section{Conclusions}

Our study of compactified QCD(adj) on the lattice has extended the present knowledge of the phase diagram of this class of theories and our previous analysis in several aspects. We have focused in particular on the $\mathcal{N}=1$ supersymmetric Yang-Mills theory. We have analyzed the lattice theory in the perturbative setup and determined in particular the deviations from continuum perturbation theory. Already in the perturbative calculations, there is a clear difference between the lattice and the continuum predictions for the phase diagram. From these calculations we have conjectured the phase diagram of QCD (adj) on the lattice and found good agreement with the observations in our first results \cite{Bergner:2014dua}. These first investigations have been limited by the difficult issue of renormalization of the bare parameters. 
Therefore we have chosen a fixed scale setup in the present investigations. 

We have measured a number of different observables in order to characterize the different phases that appear as a function of the compactification radius $R$. Up to small $R$ our observations are in accordance with the analytical predictions: there is no phase transition and a flattening of the effective potential of the Polyakov line is observed towards smaller $R$. Chiral symmetry remains spontaneously broken in the confined phase up to small $R$. 

At very small $R$ there is a phase with considerable deviations from the continuum predictions. The fermion discretization leads to a behaviour expected for a larger number of fermion flavours in the continuum. Decreasing $R$, a continuous extension of the deconfinement transition line towards smaller $m$ is expected for continuum SYM. On the lattice the confined phase extends instead to the large $m$ limit when $R$ tends to zero at a fixed lattice spacing. A peak in the Polyakov line susceptibility that is not related to deconfinement indicates the transition to this phase with considerable discretization effects. We have found indications that the deviation between the thermal deconfinement transition and this peak of the susceptibility increases towards the continuum limit. If this observation can be further substantiated, it would confirm the continuity of the deconfinement transition and the validity of the semiclassical analysis. The appearance of the 
lattice artefact phase can already  be understood from the perturbative analysis. It is furthermore characterized by a decreasing chiral condensate and a negative adjoint Polyakov loop.

Despite these differences between lattice and continuum in the small $R$ regime, the general prospects of the investigation of compactified QCD (adj) on the lattice are quite good. Even the phase with strong deviations between lattice and continuum the theory remains confined. The region where the properties of the theory are most similar to the continuum expectation is close to the peak of the susceptibility. At this point the nearly flat effective potential provides an ideal playground to investigate non-perturbative semiclassical effects. 

Already the case of the $N_f=2$ QCD(adj) is quite difficult to investigate due to the near conformal behaviour. Our observations show a deconfined intermediate phase, but a more precise analysis is required to investigate the large volume and small mass regime of this theory. The interest to this theory is also related to the determination of possible candidates for a walking technicolour scenario.

In addition our investigations are related to the considerations of supersymmetric Yang-Mills theory compactified to lower space times. In some of these investigations the compactification is done on the lattice setting $N_t=1$ in the compacitified direction. From the perspective of our investigations, this might lead to sizable deviations from the theory that is compactified in the continuum. For example a three dimensional SYM theory obtained from simulations at $R/a=1$ would be in the confined lattice perturbative phase that we have observed. We plan to investigate whether this leads to further effects when the theory is compactified down to two dimensions.

\acknowledgments
The authors gratefully acknowledge the Gauss Centre for Supercomputing e.V. (\url{www.gauss-centre.eu}) for funding this project by providing computing time on the GCS Supercomputer SuperMUC at Leibniz Supercomputing Centre (LRZ, \url{www.lrz.de}).  We thank Tin Sulejmanpasic for collaboration in early stages of this work,   Misha Shifman and Aleksey Cherman for many discussions on Abelian vs. non-Abelian confinement. 
We thank Gernot Münster, Istvan Montvay, and Philipp Scior for helpful comments on the draft.
G.B. acknowledges support from the Deutsche Forschungsgemeinschaft (DFG) Grant No. BE 5942/2-1. 
M. \" U.   acknowledges support from   U.S. Department of Energy, Office of Science, Office of Nuclear Physics under  
Award Number DE-FG02-03ER41260.

\end{document}